\documentclass[journal]{new-aiaa}
\usepackage[utf8]{inputenc}
\usepackage{textcomp}

\usepackage{graphicx}
\usepackage{amsmath}
\usepackage[version=4]{mhchem}
\usepackage{siunitx}
\usepackage{longtable,tabularx}
\usepackage{graphicx}
\usepackage{subcaption}
\usepackage{algorithm}
\usepackage{algpseudocode}

\newcommand{\etal}[0]{\emph{et al.}}

\title{Diffusion Model Driven Airfoil Design: \\From Geometry Encoding to Practical Applications \footnote{This is the accepted manuscript of an article published in AIAA Journal. Final published version available at: https://doi.org/10.2514/1.J066962}}

\author{
    Yingfan Geng \footnote{Research Student, Department of Mechanical Engineering.}, 
    Jinhong Wang \footnote{Research Associate, Department of Mechanical Engineering; Equal first author.}, 
    and Teng Cao \footnote{Assistant Professor, Department of Mechanical Engineering; t.cao@imperial.ac.uk (Corresponding Author).}}

\affil{Imperial College London, London, England SW7 2AZ, United Kingdom}
\begin{document}
\maketitle

\begin{abstract}
The diffusion model, the state-of-the-art generative machine-learning architecture, has shown promising results in airfoil inverse designs. In this study, we implement and trained a series of diffusion models on three different airfoil geometry data encoding formats -- principal component weights, ordered $x$-$y$ coordinates, and 2D signed distance functions (SDF) -- to generate 2D airfoils. By systematically comparing the performance of diffusion models trained on different data structures, it is found that for 2D airfoil design problems, the diffusion model performs the best when directly trained with coordinates. Training with latent space (principal component weights in this study) limits the model's design freedom and decreases the training effectiveness. Although the 2D SDF data appear to result in the least performing model, it proves its feasibility in aerodynamic shape generation, paving the way toward 3D problems where SDF is more favored. This study also investigated deploying the diffusion model in practical engineering applications. A multi-target optimization procedure is proposed based on the stochastic nature of the diffusion process, which drastically simplifies the procedure compared to conventional methods. The extrapolation performance of the model is also investigated by tasking the model with both aerodynamic and flow condition labels that are extrapolated beyond the training set boundaries. 
\end{abstract}

\newpage
\section*{Nomenclature}

{\renewcommand\arraystretch{1.0}
\noindent\begin{longtable*}{@{}l @{\quad=\quad} l@{}}

\multicolumn{2}{@{}l}{\textit{Roman letters}}\\
$C_L$& lift coefficient \\
$C_D$& drag coefficient \\
$C_M$& pitch moment coefficient \\
$C_P$ & pressure coefficient \\
$D$  & neural network \\
$Ma$ & Mach number \\
$Re$ & Reynolds number \\
$\mathbf{x}$ & data \\

\multicolumn{2}{@{}l}{\textit{Abbreviations}}\\
AOA & angle of attack \\
CFD & computational fluid dynamics\\
CST & class-shape transformation\\
DDPM & denoising diffusion probabilistic model\\
EDM & elucidating diffusion model \\
GAN & generative adversarial network \\
PCA & principal component analysis \\
RMSE & root-mean-square error \\
SDF & signed distance function \\
UIUC & University of Illinois Urbana-Champaign\\
VAE & variational auto-encoder\\

\multicolumn{2}{@{}l}{\textit{Greek letters}}\\
$\varepsilon$ & tolerance \\
$\sigma$ & noise level \\

\end{longtable*}}
\setcounter{table}{0}
\newpage


\section{Introduction}
\label{intro section}
\lettrine{A}{}erodynamic optimization plays a critical role across aerospace applications, with airfoil design as a central example because it determines the aerodynamic efficiency and, ultimately, the economic performance of an aircraft. Airfoil design, which is essentially finding a geometry that meets specified performance targets under given operating conditions, is a challenging inverse problem due to non-linear flow physics, multi-objective trade-offs, and strict geometric and operational constraints.

Traditional airfoil design workflows follow iterative loops together with geometry parameterizations. Initial design begins with a selection of parameterization (e.g., NACA families or class-shape transformation (CST) function~\cite{Jacobs1933}), and empirically derived guidelines are followed. The initial design concepts were then validated with computational fluid dynamics (CFD) or wind-tunnel testing for more accurate performance metrics~\cite{Martins2022}. The geometry parameters are then optimized until all requirements are met. Although the geometry parameterization reduces the computational costs, it also limits the admissible design space and can lead to suboptimal outcomes. This motivates the development of high-dimensional inverse-design methods for broader, more efficient exploration without the constraints induced by geometry parameterization. A typical example of this is the free-form deformation optimization~\cite{Samareh2004}, where the 3D geometry mesh is directly optimized using the adjoint gradient method. 

With the significant development of machine learning in recent years, both machine learning-based forward surrogate and inverse design models have emerged. Forward surrogate models~\cite{Sharpe2025} use trained neural networks as a replacement for physical simulation tools (e.g., XFOIL~\cite{Drela1989} and CFD) to reduce the computational resources required during design iterations. On the other hand, machine learning-assisted inverse design methods have demonstrated the game-changing design solution accuracy and diversity in many engineering design problems, such as thermal~\cite{Ignuta2026} and airfoil designs~\cite{Deng2026}. The trained generative models can directly output solutions that match a series of condition inputs. Typical example architectures are the variational auto-encoder (VAE)~\cite{Wang2022}, generative adversarial network (GAN)~\cite{Tan2022, Yonekura2021}, and the invertible neural networks~\cite{Glaws2022}. Furthermore, Lei \etal~\cite{Lei2021} also used the contrastive language-image pre-training (CLIP) architecture to achieve airfoil generation with natural language as inputs. Compared to traditional methods, generative models greatly expand the design space and simplify the optimization procedure. The stochastic nature of the generative models aligns naturally with the one-to-many mapping characteristic of inverse design problems: multiple distinct geometries can satisfy the same target performance metric. Unlike deterministic inverse design methods, which produce a single solution, the generative models inherently generate diverse designs that meet the specified targets.

More recently, diffusion probabilistic model~\cite{Ho2020}, inspired by non-equilibrium thermodynamics, has emerged as a powerful class of generative models and has demonstrated state-of-the-art performance across a wide range of applications, including image generation~\cite{Dhariwal2021} and topology optimization~\cite{Maze2022}. Compared with VAE, diffusion models do not rely on learning a heavily compressed latent representation for generation. Consequently, it avoids the information bottleneck inherent to VAEs, which can lead to overly smooth reconstructions and the loss of fine geometric features~\cite{Vivekananthan2024}. In the context of airfoil design, this means the diffusion model can better capture the geometry details of the airfoil, which are critical to the aerodynamic performance. 
Compared to GAN, which was previously received as the most advanced generative model architecture for its high generation quality and diversity, the diffusion model uses a likelihood-based objective function, presenting a more stable behavior while maintaining excellent sampling quality and diversity. The ease of training of the diffusion model also means no major hyperparameter tuning is required, making it ideal to be the common backbone for comparing different geometry encodings in this study. From a wider perspective, the generative diffusion model has been used in other engineering applications beyond inverse design problems, such as stochastic model updating~\cite{Wang2026a}, where a posterior distribution of the uncertain model parameters is generated from the diffusion model following a series of simulation/experimental data inputs. This study also shows the potential of conditional diffusion models' one-to-many mapping characteristic in general engineering inverse problems. Despite these advantages, the application of diffusion models to airfoil design generations remains relatively unexplored. 

A common strategy in the literature is to train diffusion models in a latent space to reduce data dimensionality and promote smoothness of the generated shapes. Diniz and Fuge~\cite{Diniz2024} proposed a conditional diffusion framework that optimizes airfoil geometry under specified flow and aerodynamic conditions, conditioning on Mach number ($Ma$), Reynolds number ($Re$), and lift coefficient ($C_L$). Their approach first encodes geometry via a learned Variational Auto-Encoder (VAE), then samples airfoils with diffusion to satisfy target conditions; the objective is to minimize the drag coefficient ($C_D$) subject to geometric constraints. Wei~\etal~\cite{Wei2024} followed a similar latent-space paradigm and compared diffusion with traditional GAN baselines, reporting advantages for diffusion in airfoil generation fidelity and controllability. Beyond learned latent spaces, several works parameterize the airfoil explicitly via classical curve families or statistical bases: Wen~\etal~\cite{Wen2026} employed CST and Bézier curves, while Zhuang~\etal~\cite{Zhuang2025} used principal component analysis (PCA) to represent geometry via principal component weights. Deng~\etal~\cite{Deng2026} further explored conditioning via an intermediate physical representation by diffusing the pressure-coefficient curve $C_P$ (with a seven-parameter characterization), followed by a learned mapping from $C_P$ to the corresponding airfoil shape; this retains more aerodynamic structure than generic latent spaces. Furthermore, Lin~\etal~\cite{Lin2025} and Wu~\etal~\cite{Wu2025} investigated 3D airfoil and compressor blade generations, where geometries are also encoded via parametrized latent space. While latent-space diffusion simplifies training and enforces smoothness~\cite{Wagenaar2024}, it is effectively equivalent to geometry parameterization that was used in the traditional methods as discussed previously, and hence will potentially constrain the admissible design space and limit the exploration of novel, unconventional airfoil configurations. In contrast, direct diffusion on explicit geometry representations offers greater design freedom but poses additional modeling and stability challenges. In generic 2D/3D shape generation, researchers have explored multiple explicit formats, including point sets, point clouds~\cite{Luo2021}, voxel grids~\cite{Smith2017}, and signed distance fields (SDFs)~\cite{Park2025}. Different geometry encodings define distinct spaces for the diffusion process, ranging from linear subspaces to non-linear embedded manifolds and spatial fields, each imposing different geometric constraints and approximations of the underlying shape manifold. Within airfoil generation tasks, direct coordinate diffusion (i.e., ordered contour coordinates) has seen limited trials and mixed results~\cite{Graves2024}, and there is a lack of systematic comparisons across multiple geometry encoding methods (latent vs. explicit) under a common diffusion backbone. 

Methodologically, most airfoil studies to date adopt the original denoising diffusion probabilistic model (DDPM) \cite{Ho2020}. However, the elucidated diffusion framework (EDM)~\cite{Karras2022} has demonstrated better performances. Unlike DDPM, which is typically formulated using discrete diffusion time steps and predefined noise schedules, EDM employs a continuous noise-level parameterization and introduces optimized design choices for network preconditioning, loss weighting, noise scheduling, and numerical sampling. These modifications have been shown to improve training stability, sample quality, and sampling efficiency across a range of generative modeling tasks~\cite{Zhu2025}. Furthermore, airfoil geometries are represented as continuous-valued signals, either as coordinate vectors, PCA coefficients, or signed-distance fields. The continuous noise formulation of EDM is therefore well-suited to modeling such geometric data. Despite being promising, its application in engineering design practices still remains underexplored.


The current literature reveals three key gaps. First, there is no systematic evaluation of diffusion performance across diverse geometry encoding methods (e.g., PCA latent space, ordered coordinates, and SDF grids) within a unified framework. Second, conditioning across broad ranges of angle of attack ($AOA$), $Re$, and $Ma$ is limited with most studies rely on single-condition labels or narrow multi-condition spans. Third, state-of-the-art diffusion samplers like EDM have not been systematically examined for airfoil inverse design, despite their efficiency and stability benefits. Addressing these gaps can clarify trade-offs among representations, improve robustness across operating conditions, and accelerate practical adoption in engineering design workflows.

To address these gaps, we present a unified, conditional diffusion model based generative framework for two-dimensional airfoil inverse design. In this study, three geometry encoding methodologies are systematically evaluated -- principal component weights (PCA-based), ordered coordinates, and 2D SDF -- within an elucidated diffusion model backbone. The model jointly conditions on $AOA$, $Re$, and $Ma$ across broad ranges, demonstrating strong accuracy, robust constraint satisfaction, and deployment-ready workflows. By consolidating insights in 2D generation under realistic operating constraints, this work establishes a principled foundation for extending diffusion-based design to complex three-dimensional aerodynamic geometries.

\section{Methodology}
\begin{figure}[ht]
    \centering
    \includegraphics[width=\textwidth]{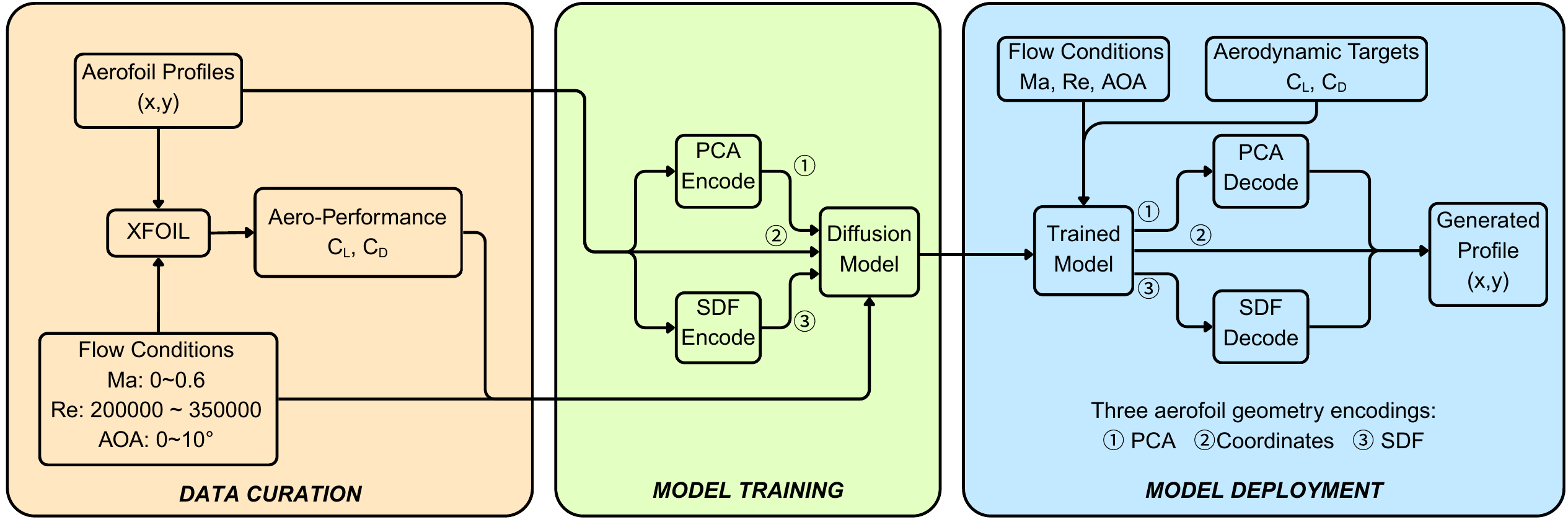}
    \caption{Workflow of Diffusion Model Training and Deployment}
    \label{workflow_diagram}
\end{figure}

In this section, the workflow of the diffusion model implementation and development is introduced, as illustrated in Fig.~\ref{workflow_diagram}.

\subsection{Data Curation and Preprocessing}

\subsubsection{Airfoil Database and Label Generation}
\label{Airfoil Database and Label Generation section}

\begin{figure}[!ht]
    \centering
    \begin{subfigure}[t]{0.6\textwidth}
        \centering
        \includegraphics[width=\textwidth]{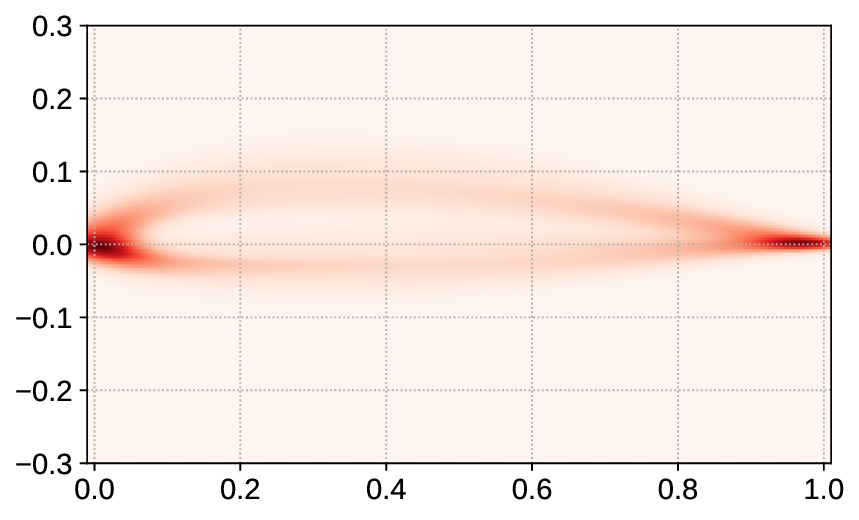}
        \caption{Airfoil profile 2D kernel density estimation, with normalised coordinates by chord length}
        \label{dataset profile distribution plot}
    \end{subfigure}%
    \\
    \begin{subfigure}[t]{0.6\textwidth}
        \centering
        \includegraphics[width=\textwidth]{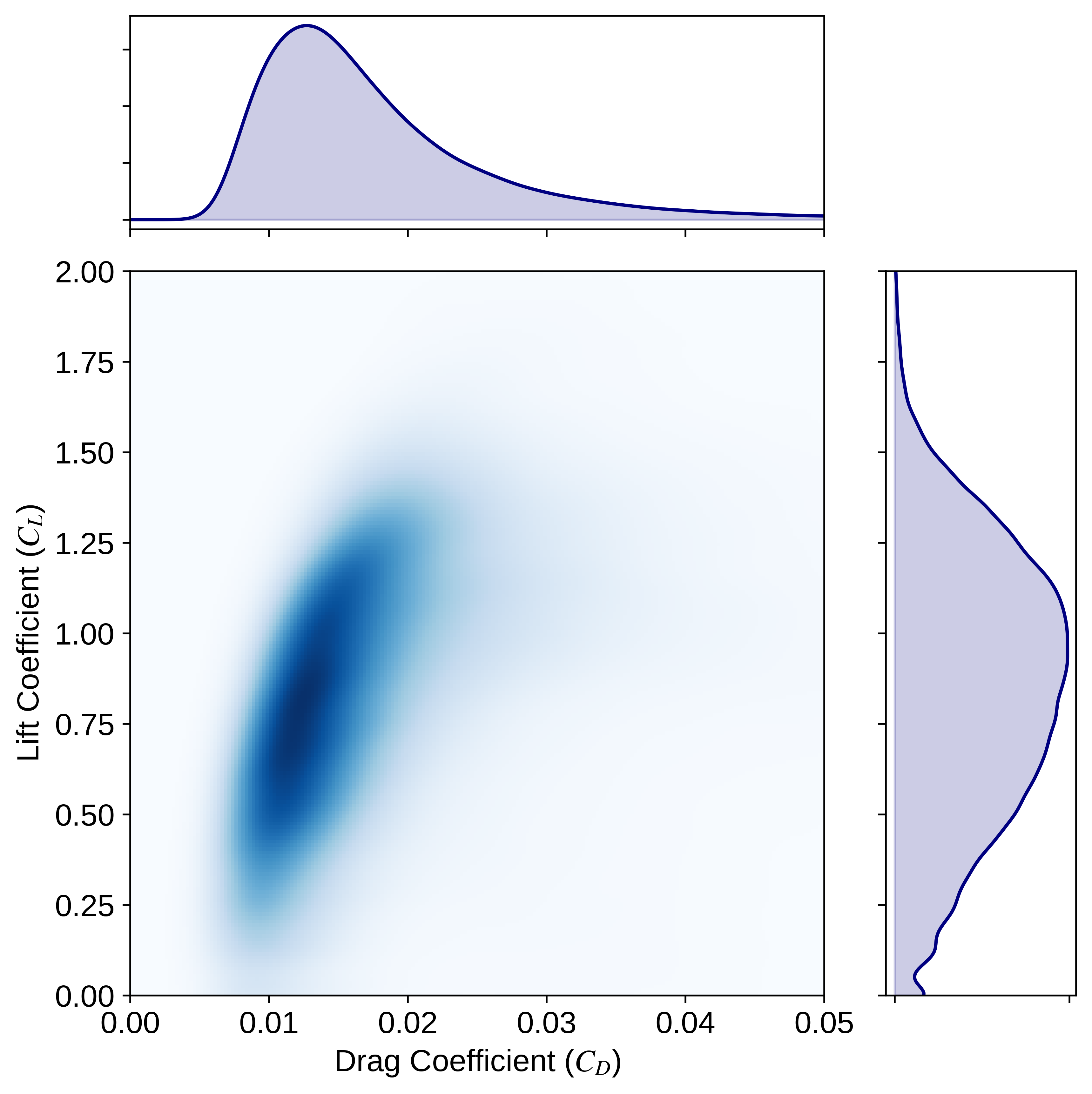}
        \caption{Aerodynamic Performance Distribution}
        \label{dataset aerodynamic distribution plot}
    \end{subfigure}

    \caption{The Diffusion Model Dataset Distribution}
    \label{dataset distribution}
\end{figure}

Airfoil geometries that were used in this study are from the UIUC airfoil database~\cite{UIUC}. This database contains approximately 1500 airfoil profiles from various design families such as the NACA and Eppler series, for various applications. After excluding the designs with non-closed trailing edge, there are in total 1474 profiles to form the model's training dataset. The airfoil chord is non-dimensionalized such that the leading and trailing edges are located at \((0,0)\) and \((1,0)\), respectively. The distribution of the airfoil profiles are illustrated as kernel density estimates (KDE) plot, shown in Fig.~\ref{dataset profile distribution plot} for reference, which reflects the mean airfoil shape in the dataset. KDE is a statistical method to estimate the probability distribution of a given set of data. Its mathematical formulation is shown in Eq.~\eqref{KDE}. Here, $\hat{p}$ is the estimated probability, $N=256$ as the number of points per airfoil, $h$ is the bandwidth, $d = 2$ as airfoil shapes are in 2D, $K$ is the Gaussian kernel function, $\mathrm{x}$ is the coordinates in the domain, and $\mathrm{x}_i$ is the airfoil geometries. 
\begin{equation}
    \label{KDE}
    \hat{p}_{\mathrm{x}} = \frac{1}{Nh^d} \sum_{i=1}^N K\left(\frac{\mathrm{x} - \mathrm{x}_i}{h} \right)
\end{equation}

XFOIL, a 2D airfoil analysis tool based on the panel method~\cite{Chen2015}, is used as the surrogate simulation model to create the aerodynamic labels~\cite{Drela1989}. Although providing fast computational speed, XFOIL also achieves comparable physical infidelities as CFD~\cite{Morgado2016}. In practice, the workflow proposed in this study can be directly implemented in studies with other data (CFD or experiments), and XFOIL is used here as a fast and consistent data source for proof of concept. XFOIL's interpolation tool is also used to smooth the profile shape and ensure an equal number of coordinate points for each profile. In this study, 256 points per profile are maintained. Flow conditions are sampled on a uniform grid: Reynolds number from $2\times10^5$ to $3.5\times10^5$ in increments of $5\times10^4$; Mach number from $0$ to $0.6$ in increments of $0.1$; and angle of attack from $0^\circ$ to $10^\circ$ in steps of $1^\circ$. All endpoints are inclusive. 

The corresponding $C_L$ and $C_D$ for different airfoils are then calculated using XFOIL, with their distribution shown in Fig.~\ref{dataset aerodynamic distribution plot}. The specified flow conditions ($Ma$, $Re$, and $AOA$) and the aerodynamic performance ($C_L$ and $C_D$) form the condition label of the diffusion model. For flow-condition combinations under which an airfoil is unlikely to operate properly (e.g., post-stall at high $AOA$), XFOIL may fail to converge and return no solution. These cases are flagged as non-convergent and excluded from the dataset. To reduce the impact of outliers, samples outside the range of $C_L \in [0,\,2.0]$ or $C_D \in [0,\,0.05]$ are excluded (endpoints inclusive).The final dataset contains 372{,}924 data entries.

\subsubsection{Geometry Representation}
\label{geometry representation subsubsection}

\begin{figure}[!ht]
    \centering
    \begin{subfigure}[t]{0.7\textwidth}
        \centering
        \includegraphics[width=\textwidth]{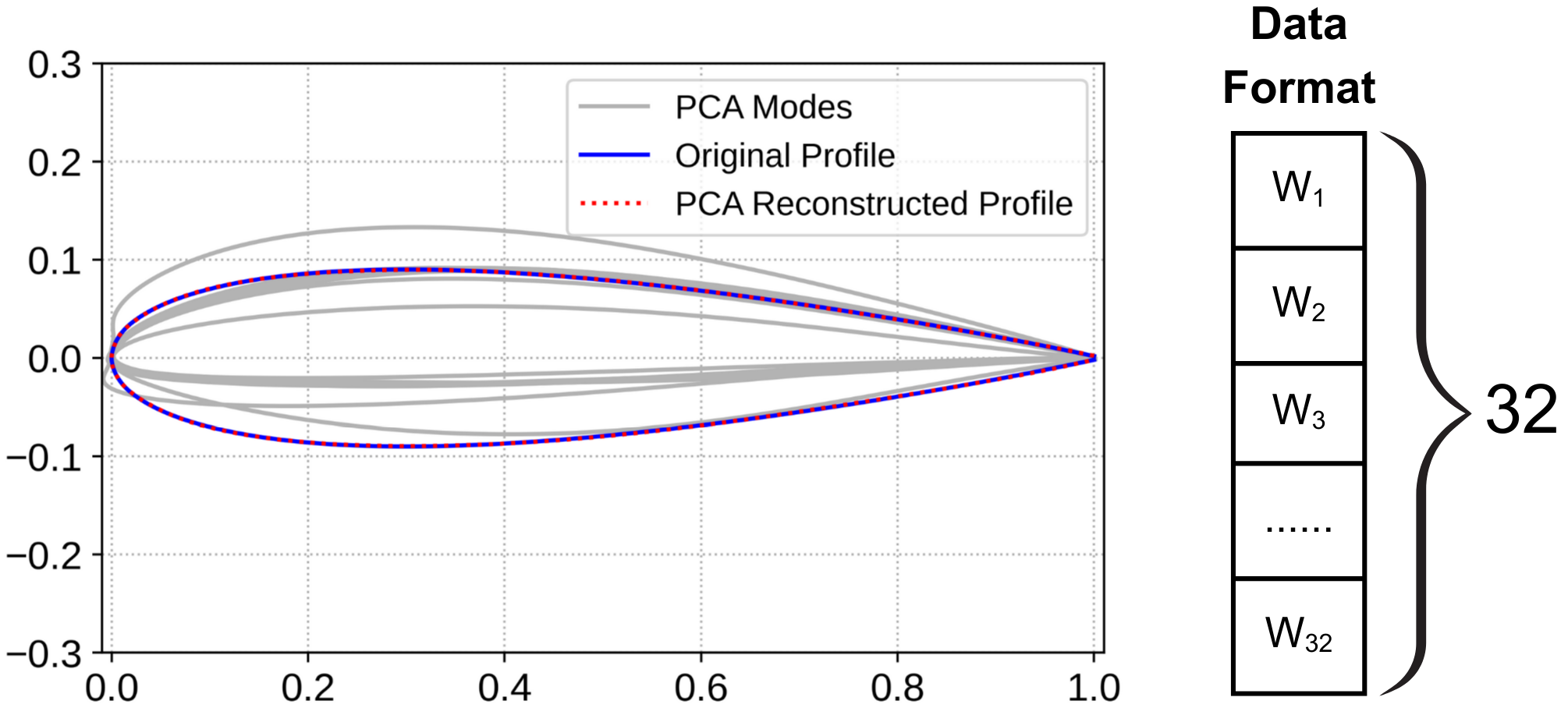}
        \caption{Principal Component Weights}
        \label{PCA dataformat}
    \end{subfigure}
    \hfill
    \begin{subfigure}[t]{0.7\textwidth}
        \centering
        \includegraphics[width=\textwidth]{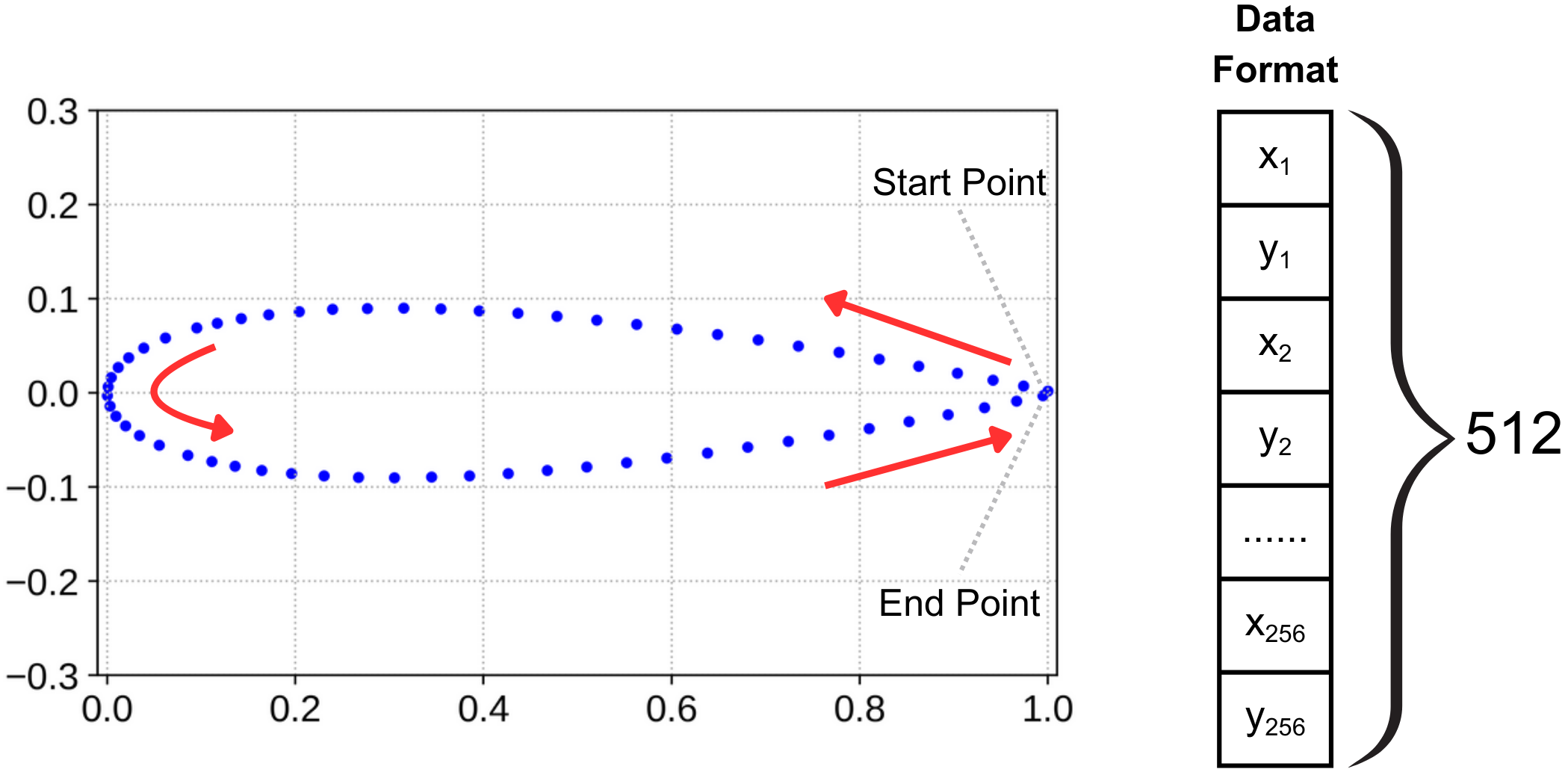}
        \caption{Ordered $x$-$y$ Coordinate}
        \label{coordinate data format}
    \end{subfigure}
    \hfill
    \begin{subfigure}[t]{0.8\textwidth}
        \centering
        \includegraphics[width=\textwidth]{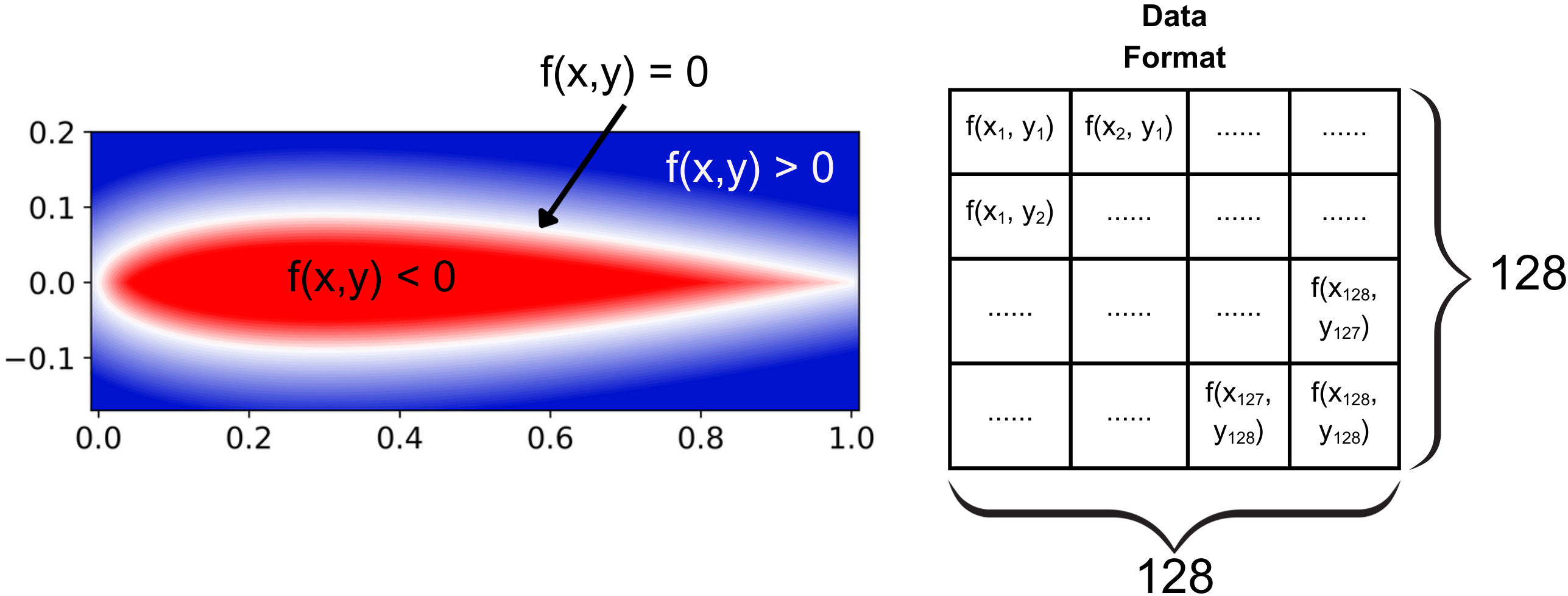}
        \caption{Signed Distance Function (SDF)}
        \label{SDF dataformat}
    \end{subfigure}
    
    \caption{Different Encoding Methods of the Airfoil Shape with Corresponding Data Format}
    \label{data formats}
\end{figure}

Conceptually speaking, there are two main streams of how the geometry information of an airfoil can be encoded -- directly and indirectly. As introduced in Section~\ref{intro section}, the majority of previous studies in the literature have opted for the former approach to parametrize the airfoils into latent space before training the diffusion model. Although doing so reduces the data dimensionality as well as the training complexity, it also limits the model's design freedom. In contrast, using explicit geometry encodings, such as point coordinates, provides much higher design freedoms to the model; however, it may also be more challenging for the model to learn. Compared to latent-space encodings which ensure geometry smoothness, the explicit encodings may suffer from the loss of geometry smoothness. To compare the influence of geometry encoding methods on the model's performance, in this study, we present a systematic study of the effect of geometry encodings under the same diffusion model architecture. Three different airfoil geometry representation methods are used: (i) principal component weights, as a representation of the indirect parameterization-based encoding methods used in the literature; (ii) ordered \(x\)-\(y\) coordinates, as the most straightforward approach for explicit geometric encodings; and (iii) signed distance function, a another explicit encoding methods, which is used here as a proof of concept, paving the way towards further extension to 3D airfoil designs, as SDF has been considered to be more scalable to 3D in the generic object generation tasks~\cite{Chou2023}. These three encoding approaches are shown in Fig.~\ref{data formats}.

\begin{figure}[ht]
    \centering
    \includegraphics[width=8.2cm]{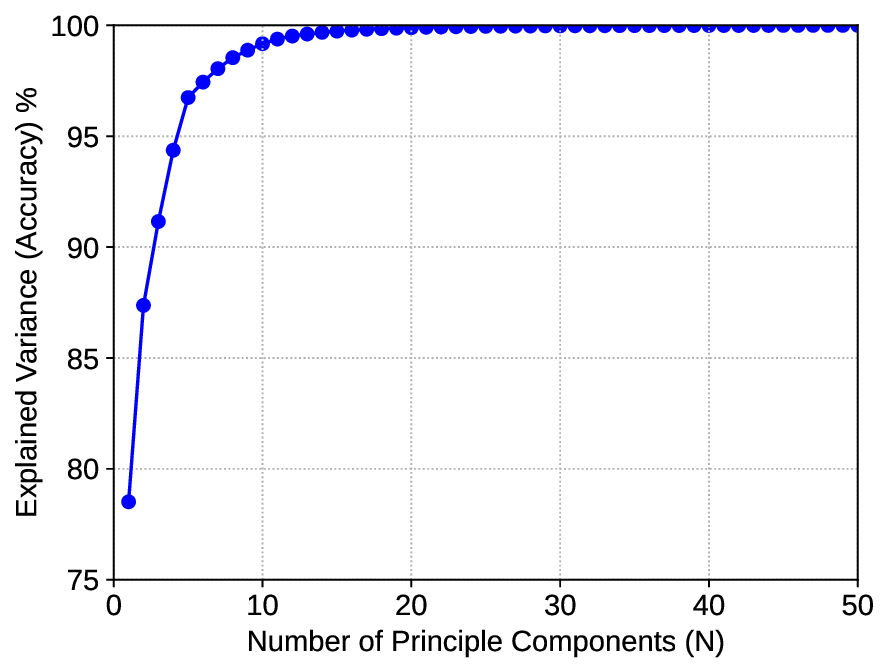}
    \caption{PCA Accuracy}
    \label{pca_accuracy}
\end{figure}

\paragraph{Principal Component Weights}
Principal component analysis is a statistical tool that maps a complex dataset to a lower dimension~\cite{Shlens2014}. It identifies the principal axes along which the data exhibit the largest variance, corresponding to the dominant modes of the dataset. In this present study, by conducting PCA, the data are reaped from $\textbf{X} \in \mathbb{R}^{256}$ to $\textbf{X} \in \mathbb{R}^{N}$ where $N$ is the number of principal components (PCs). Increasing $N$ improves the accuracy of the PCA representation of the airfoil data, which is characterized by the explained variance ratio, defined as Eq.~\eqref{pca explained variance}, 

\begin{equation}
\label{pca explained variance}
    \text{Explained Variance Ratio} = \dfrac{\sum_{i=1}^{N} \lambda_i}{\sum_{j=1}^{256} \lambda_j}
\end{equation}

where $\lambda$ is the eigenvalue of the covariance matrix of the corresponding PC. The accuracy of PCA with different $N$ is shown in Fig.~\ref{pca_accuracy}. In this study, $N=32$ is chosen, which retains 99.99\,\% of the variance to ensure that there is only a negligible amount of information loss during the PCA process. As a result, the PCA-based data encoding produces an array of 32 real numbers, as shown in Fig.~\ref{PCA dataformat}, each of representing the weight of the corresponding principal component (shown as gray curves). The original profile (blue curve) can then be reconstructed by a linear combination of these 32 principal components using the corresponding principal component weights (reconstructed profile shown as red curves). The PCA weights are fitted after coordinate normalization for consistency.

\paragraph{Ordered coordinate} 

The second approach takes $x$-$y$ coordinates directly from the dataset with no additional encoding. The coordinates are arranged in a specific order: counter-clockwise starting from the trailing edge to the leading edge, and returning to the trailing edge, as illustrated in Fig.~\ref{coordinate data format}.
This is consistent with the default arrangement from the UIUC database~\cite{UIUC}, and matches the required profile data format by XFOIL~\cite{Drela1989}. This also enhances the spatial information, which benefits the subsequent training of the diffusion model. 

\paragraph{Signed distance function (SDF)}

Signed distance function, also referred to as a signed distance field, is a continuous function that describes the shortest distance from any point to the object boundary. Points inside the object have negative SDF values while points outside have positive values. The points on the object edge have SDF values of 0. Previous studies in the field of generic object generation has shown the advantages of using SDF compared to other data formats~\cite{Park2025}. In this study, the SDF matrix is defined with a size of $128\times 128$, as shown in Fig.~\ref{SDF dataformat}. The SDF values $f(x,y)$ are normalized to $[-1, 1]$ to retain the sign information during training. During the SDF decoding process, the zero-valued contour points are extracted. To ensure profile smoothness, quadratic interpolation is applied, and the recovered geometry points are reordered in the manner as in Fig.~\ref{coordinate data format} for XFOIL compatibility. Although this study focuses solely on 2D airfoil designs, SDF is examined as one of the geometry enconding methods, with potential for extending toward 3D airfoil designs. Compared to point cloud coordinates, which are often unstructured in 3D, the use of SDF can be more convenient as it presents a structured matrix, making it a more suitable data format for 3D object generation.

It is worth noting that different geometry encodings implicitly define different spaces in which the diffusion process is performed. PCA-based representations constrain diffusion to a linear, low-dimensional subspace that approximates the underlying shape manifold, enforcing global geometric correlations but limiting design space. Coordinate-based representations perform diffusion in a highly non-linear and curved manifold embedded in Euclidean space, where local perturbations may easily violate geometric regularity. In contrast, signed distance function representations define diffusion over a spatially continuous field with implicit geometric constraints, promoting smoother and more physically consistent intermediate samples. As a result, each encoding corresponds to a distinct approximation of the true geometry manifold, which directly influences diffusion dynamics, sample validity, and inverse design performance.

\subsubsection{Data Summary}
To ensure equal contribution of all features to the model, all condition labels ($Ma$, $Re$, $AOA$, $C_L$, and $C_D$) are normalized to a range of 0 $\sim$ 1 using minimum-maximum normalization. The dataset is then split into training, validation, and testing sets with an 8:1:1 ratio based on the data entries. The same dataset was used for all three data structures for comparison. It should be noted that partitioning at the data-entry level allows identical airfoil geometries to appear in different dataset subsets. However, each data entry in the dataset corresponds to a unique combination of operating conditions ($Ma$, $Re$, and $AOA$), aerodynamic performance targets($C_L$ and $C_D$), and geometry ($\textbf{X}$). The objective of the present study is to learn the inverse mapping from flow conditions and aerodynamic performance to airfoil geometry, rather than to assess extrapolation to completely unseen airfoil families. Consequently, the dataset split was performed at the level of individual inverse-design mappings rather than geometric profiles. Under this setting, an airfoil geometry observed during training is associated with a different set of operating conditions and performance targets than those encountered during validation or testing. Therefore, the model is evaluated on previously unseen inverse-design instances, despite partial overlap in the underlying airfoil geometries.

\subsection{Diffusion Framework}
\label{diffusion framework}

A conditional diffusion model is used as the backbone in this study. 
Conceptually, the model is a denoising predictor, that receives a noisy geometry signal together with a conditioning vector (operating/design labels) and a noise level, and returns a prediction of the corresponding clean signal. The core denoising functionality is implemented via a neural network, denoted here by \(D\), which will be introduced in Section~\ref{method network architecture subsection}. The overall implementation builds upon the implementations provided in~\cite{Zhuang2025, Karras2022}.

The mathematical formulation of the diffusion model used in this study follows the EDM framework introduced by Karras \etal~\cite{Karras2022}, which is a variant from the original DDPM model~\cite{Ho2020} with improved performance. The fundamental diffusion process is governed by the probability flow ODE and the stochastic diffusion equation (SDE), as shown in Eq.~\eqref{edm equation}~\cite{Song2021}. 
\begin{equation}
    \label{edm equation}
    \mathrm{d}\mathbf{x}_{\pm} = \underbrace{-\dot{\sigma}(t) \sigma(t) \nabla_{\mathbf{x}} \log p_t (\mathbf{x}; \sigma(t)) \mathrm{d}t}_{\text{probability flow ODE}} \pm 
    \underbrace{\beta(t) \sigma(t)^2 \nabla_{\mathbf{x}} \log p_t (\mathbf{x}; \sigma(t)) \mathrm{d}t + \sqrt{2\beta(t)} \sigma(t) \mathrm{d}\omega_t}_{\text{Langevin diffusion SDE}}
\end{equation}
The first term (deterministic probability flow ODE) guides the diffusion trajectory towards the region of higher likelihood for the clean data. The second part of Eq.~\eqref{edm equation} is the Langevin SDE, which introduces stochasticity to the diffusion process to increase the diversity of generated outputs and is emulated by using a temporary addition of Gaussian noise. 

In Eq.~\eqref{edm equation}, $t \in [0, T]$ denotes a continuous time variable of the diffusion process, $\mathbf{x}$ is the data, $p_\mathbf{x}$ is the probability distribution of the data, and $\sigma$ is the standard deviation of the signal as a representation of the noise level. The score function $\nabla_x \log p_t (x; \sigma(t))$ is defined as Eq.~\eqref{diffusion score} \cite{Song2021} and it points towards the higher data density region at a given noise level of $\sigma(t)$.
\begin{equation}
    \label{diffusion score}
    \nabla_{\mathbf{x}} \log p_t (\mathbf{x}; \sigma(t)) = \frac{D(\mathbf{x};\sigma)-\mathbf{x}}{\sigma^2}
\end{equation}

The training objective follows the Elucidating Diffusion Model formulation~\cite{Karras2022} and minimizes the expected weighted mean-squared error between the network output and the ground truth. The weighting function \(\lambda(\sigma)\) and the scaling coefficients \(c_{\text{in}}(\sigma)\), \(c_{\text{out}}(\sigma)\), \(c_{\text{skip}}(\sigma)\), and \(c_{\text{noise}}(\sigma)\) are noise-dependent hyperparameters defined according to the EDM framework. The overall objective is to minimize the expected loss given in Eq.~\eqref{loss function}.
\begin{equation}
\label{loss function}
    \mathbb{E}_{\sigma, x, n} \left[ \underbrace{\lambda(\sigma) c_{\text{out}}^2}_{\text{weight}}
    || \underbrace{F_\theta (c_{\text{in}}(\sigma)\cdot (x+n); c_{\text{noise}(\sigma)})}_{\text{neural network output}}
    - \underbrace{\frac{1}{c_{\text{out}}(\sigma)}(x - c_{\text{skip}}(\sigma) \cdot (x+n))}_{\text{ground truth}}
    ||^2_2
    \right]
\end{equation}

After training, the denoising process is explained in Algo.~\ref{edm generation code}. The initial signal $\textbf{x}_0$ is sampled from a pure Gaussian noise, and the diffusion model is applied to progressively denoise the signal until a clean sample $\textbf{x}$ is obtained. The noise level scheduled along the denoising trajectory is scheduled by $\left( \sigma_{\text{max}}^{\frac{1}{\rho}} + \frac{i}{N-1} \left( \sigma_{\text{min}}^{\frac{1}{\rho}} - \sigma_{\text{max}}^{\frac{1}{\rho}}    \right)  \right)^\rho$, where $N$ is the total number of denoising steps, $i$ denotes the current denoising step and $\rho$ is a tunable hyperparameter. It is noted that during the sampling procedures, there is an additional step of temporary noise addition (line 6 in Algo.~\ref{edm generation code}), which adds additional stochasticity to achieve greater diversity in generated designs. In addition, compared to the original DDPM sampling process~\cite{Ho2020}, the EDM architecture samples the noise with a second-order approach (lines 9 -- 12 in Algo.~\ref{edm generation code}), which reduces the amount of intermediate steps required during design generation. In this study, all hyperparameters settings are kept as default as in Ref.~\cite{Karras2022}. 

\begin{algorithm}
    \caption{Diffusion model generation}
    \label{edm generation code}
    \begin{algorithmic}[1]
    \State \textbf{procedure} Stochastic Sampler ($D (\textbf{x}; \sigma), t_{i\in \{0, ..., N \}}, \gamma_{i\in \{0, ..., N-1 \}}, S_{\text{noise}}$) 
    \State \textbf{sample} $\textbf{x}_0 \sim \mathcal{N}(\textbf{0}, t_0^2\textbf{I})$
    \For {$i\in \{ 0, ..., N-1\}$}
        \State \textbf{sample} $\epsilon_i \sim \mathcal{N}(\textbf{0}, S_{\text{noise}}^2\textbf{I})$
    
        \State $\gamma_i = \begin{cases}
            \text{min} \left( \frac{S_{\text{churn}}}{N}, \sqrt{2}-1 \right) &\text{if}~t_i \in [S_{\text{tmin}}, S_{\text{tmax}}] \\
            0 &\text{otherwise}
        \end{cases}$
        
        \State $\hat{t}_i \gets t_i + \gamma_it_i$ \Comment{add noise temporarily to emulate the stochastic diffusion}
        \State $\hat{\textbf{x}}_i \gets \textbf{x} + \sqrt{\hat{t}_i^2 - t_i^2} \epsilon_i$
        \State $\textbf{d}_i \gets (\hat{\textbf{x}}_i - D (\hat{\textbf{x}}_i; \hat{t}_i))/\hat{t}_i$
        \State $\textbf{x}_{i+1} \gets \hat{\textbf{x}}_i + (t_{i+1} - \hat{t}_i) \textbf{d}_i$ \Comment{linear Euler step for solving probabilistic flow ODE}
    
        \If {$t_{i+1} \neq 0$}
            \State $\textbf{d}_i' \gets (\textbf{x}_{i+1} - D (\textbf{x}_{i+1}; t_{i+1})) / t_{i+1}$ \Comment{Huen method correction}
            \State $\textbf{x}_{i+1} \gets \hat{\textbf{x}}_i + \frac{1}{2}(t_{i+1} - \hat{t}_i)(\textbf{d}_i + \textbf{d}_i')$
        \EndIf
    \EndFor
    \end{algorithmic}
\end{algorithm}


\subsection{Model Architecture}
\label{method network architecture subsection}
\begin{figure}[ht]
    \centering
    \includegraphics[width=\textwidth]{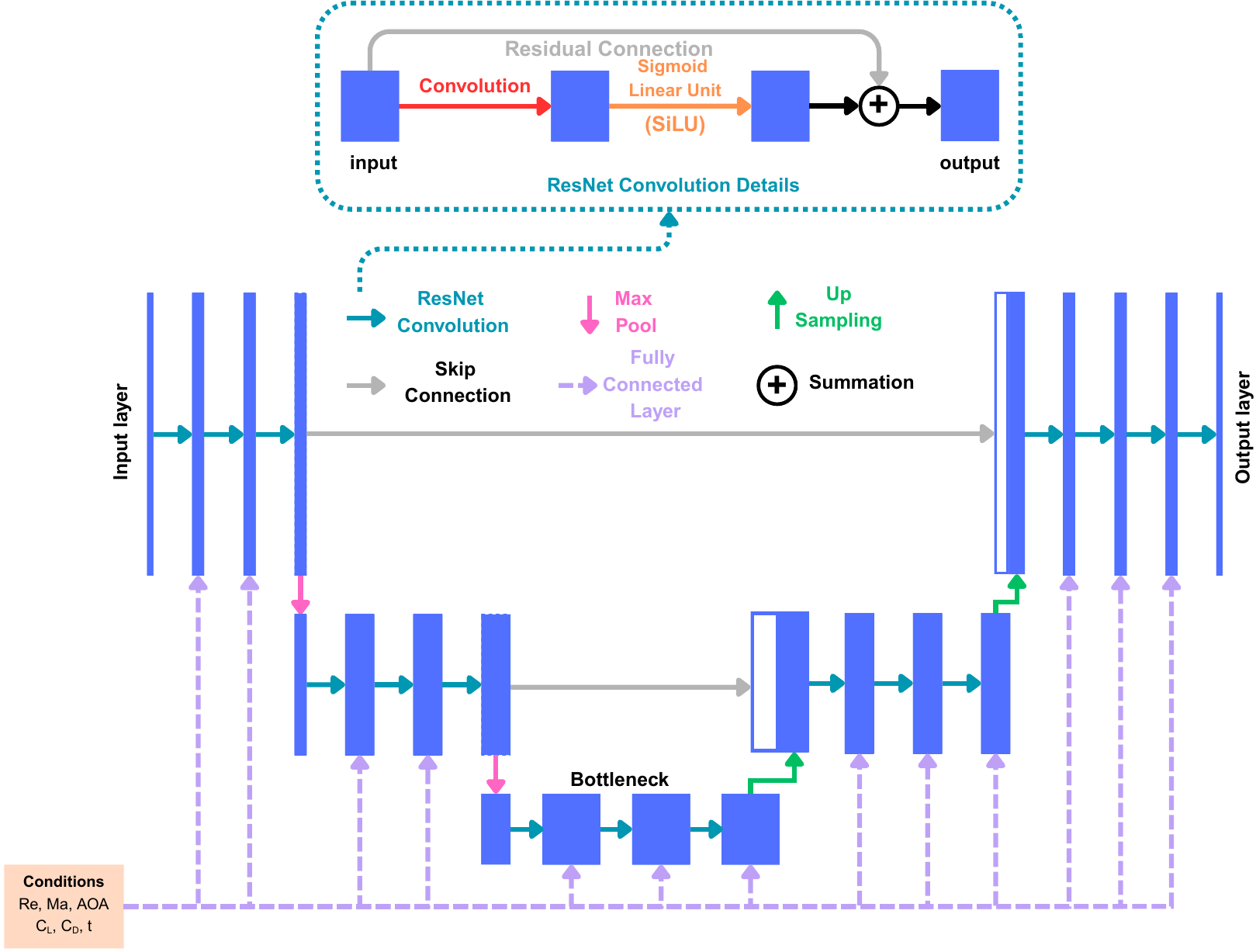}
    \caption{Neural Network Architecture: U-Net with ResNet blocks}
    \label{nn architecture}
\end{figure}

As introduced in Section~\ref{diffusion framework}, a neural network $D$ is trained as the core of the denoising process. In this present study, U-Net~\cite{Ronneberger2015} is implemented as the noise predictor. Because of its symmetric architecture with the same input and output dimensions, U-Net is the most commonly adopted network architecture in diffusion models~\cite{Ho2020, Liu2023, Wen2026}. In addition, residual networks (ResNet~\cite{He2015}) are adopted as the basic convolutional building blocks. The overall architecture is illustrated in Fig.~\ref{nn architecture}. 

The network architecture is identical for the three input data formats introduced in Section~\ref{geometry representation subsubsection}, with five ResNet blocks per stage and 5 stages in total. First, the input data are mapped to a 64-channel feature map via a lifting convolution layer. Then the data are progressively downsampled via the encoder, which consists of a series of residual convolution blocks and maximum-pooling layers. For each ResNet block, the spatial resolution and number of feature channels remain unchanged, while each maximum-pooling layer halves the spatial resolution and doubles the feature channels. 

After the bottleneck, as a symmetric architecture, the data are upsampled to gradually restore their original resolution while reducing the feature channels. For each upsampling layer, the data resolution is doubled while the feature channel halves. The upsampled features are then concatenated with the corresponding features from the encoder via skip connections, enhancing the fusion of low- and high-resolution information and helping to recover details lost during downsampling.

The condition parameters, including the sampling time step $t$ and flow and aerodynamic labels ($Ma$, $Re$, $AOA$, $C_L$, $C_D$), are added to each ResNet block in the U-Net via a fully connected linear perceptron layer, ensuring that the model is consistently constrained by the conditions. The conditions are embedded using a sinusoidal position embedding. 

For PCA-based and coordinates-based diffusion, the model uses 1D convolutional layers with kernel size $3\times1$, while 2D convolutional layers with kernel size $3\times3$ are used for SDF-based diffusion. Padding of size 2 is applied in all cases. 

The training of PCA- and coordinate-based models are carried out on an Nvidia RTX 5080, while the SDF-based model is trained on A100 due to its larger video random access memory (VRAM) requirement. The details for each of the three data formats are shown in Table~\ref{training_details_table}. All models were trained using the AdamW optimizer~\cite{Loshchilov2019} with a learning rate of 0.0001, and a cosine annealing learning rate scheduler is adopted as a standard deep learning practice. All hyperparameters are kept identical for all three data structures.

\begin{table}[hbt!]
    \caption{Summary of Diffusion Model Training Details with the Three Data Formats}
    \label{training_details_table}
    \centering
    \begin{tabular}{ccccc}
    \hline
    \textbf{Data Format}& \textbf{Model Param No.}& \textbf{Batch Size}& \textbf{Epochs} & \textbf{Training Time (hrs)}\\
    \hline
    PCA & 61,785,601 & 64 & 100 & $\sim$12 \\
    Coordinates& 61,785,986 & 64 & 100 & $\sim$24\\
    SDF& 166,480,129 & 64 & 100 & $\sim$72\\
    \hline
    \end{tabular}
\end{table}

\subsection{Model Deployment and Evaluation}
\label{method model deployment and applications section}

\subsubsection{Geometry Generation}
The trained diffusion model can then be used to generate airfoils using Algo.~\ref{edm generation code} with specified flow conditions ($Ma$, $Re$, and $AOA$) and aerodynamic targets ($C_L$ and $C_D$). The actual $C_L$ and $C_D$ values of the generated profiles are validated using XFOIL, and ideally, the predicted values should match the target values. 

The diffusion model is trained to approximate the conditional distribution of airfoil geometries associated with a prescribed aerodynamic target. Under the manifold hypothesis, physically feasible airfoils occupy a low-dimensional subset of the high-dimensional representation space of geometries. Therefore, for a given design target, the corresponding solutions are assumed to form a conditional distribution supported on a subset of the airfoil geometry manifold.
However, due to finite training data and model approximation errors, the learned distribution generally differs from the true conditional distribution.
Furthermore, given that the aerodynamic performance of an airfoil tends to be very sensitive to its geometry, the model-generated geometries may not always achieve the desired $C_L$ and $C_D$ values. Hence, multiple trials may be required to satisfy the target values within a specified tolerance $\varepsilon$. Therefore, the single-target geometry generation workflow used in this paper is illustrated by Fig.~\ref{single-target generation flow chart}. As the tolerance $\varepsilon$ becomes tighter, the design difficulty increases, requiring more design generation trials from the model. To retain the diffusion model stochasticity for design solution diversity, the single-target design process represented here is an open-loop workflow. If the design generated does not satisfy the target within tolerances, the diffusion model is initialized again with a different randomly sampled initial Gaussian noise. The design generation resulting from the previous trial will not affect the next trial. In other words, the design generations are independent events. 

\subsubsection{Performance Evaluation}

To quantify the model error, the root-mean-square error (RMSE) of the generated airfoil $C_L$ and $C_D$ values is also computed after the single-target generation process, as defined in Eq.~\eqref{RMSE}, where $N_\text{test}$ is the number of testing samples.

\begin{equation}
    \label{RMSE}
    \textrm{RMSE}_{C_L} = \sqrt{\frac{1}{N_{\text{test}}} \sum_{i=1}^{N_{\text{test}}} \left( \frac{C_{L,i}^{\text{actual}} - C_{L,i}^{\text{target}}}{C_{L,i}^{\text{target}}} \right)^2}~~~~~~~~~~ 
    \textrm{RMSE}_{C_D} = \sqrt{\frac{1}{N_{\text{test}}} \sum_{i=1}^{N_{\text{test}}} \left( \frac{C_{D,i}^{\text{actual}} - C_{D,i}^{\text{target}}}{C_{D,i}^{\text{target}}} \right)^2}
\end{equation}

\begin{figure}[t]
    \centering
    \begin{subfigure}[t]{0.49\textwidth}
    \centering
    \includegraphics[width=0.7\textwidth]{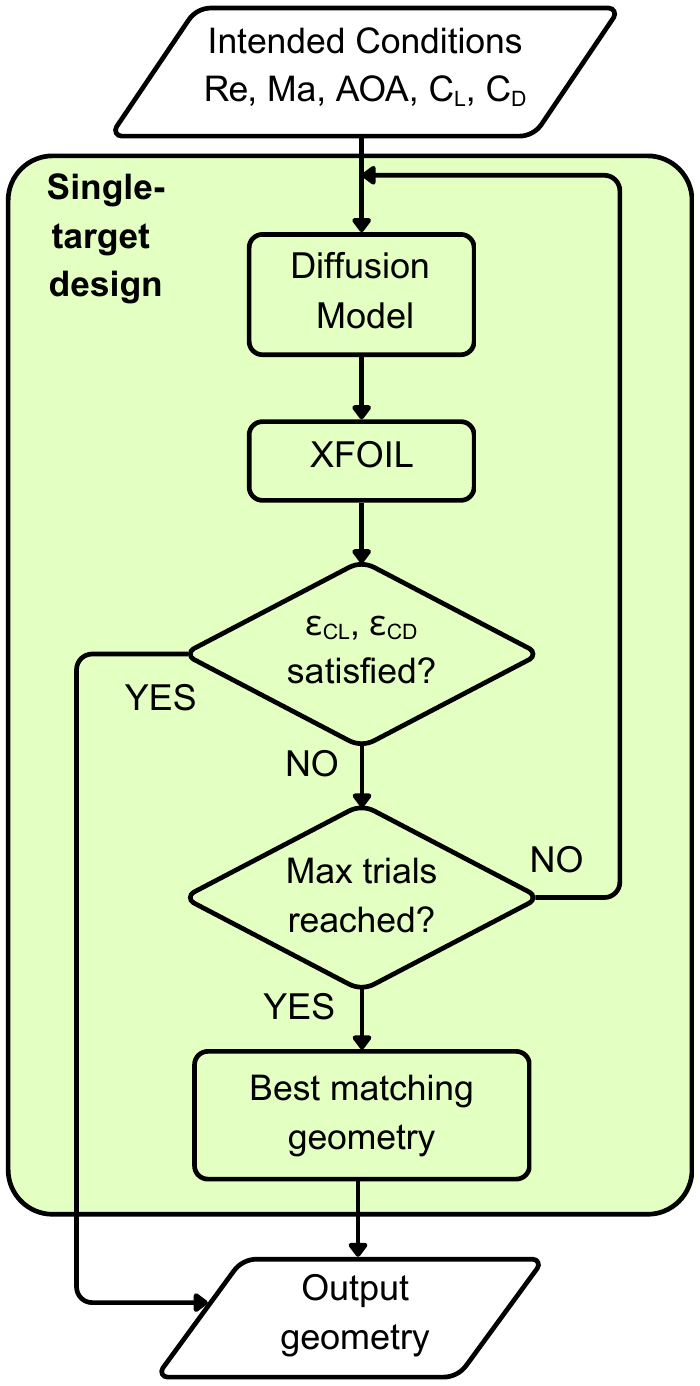}
    \caption{Single-Target Generation}
    \label{single-target generation flow chart}
    \end{subfigure}
    \hfill
    \begin{subfigure}[t]{0.49\textwidth}
    \centering
    \includegraphics[width=0.835\textwidth]{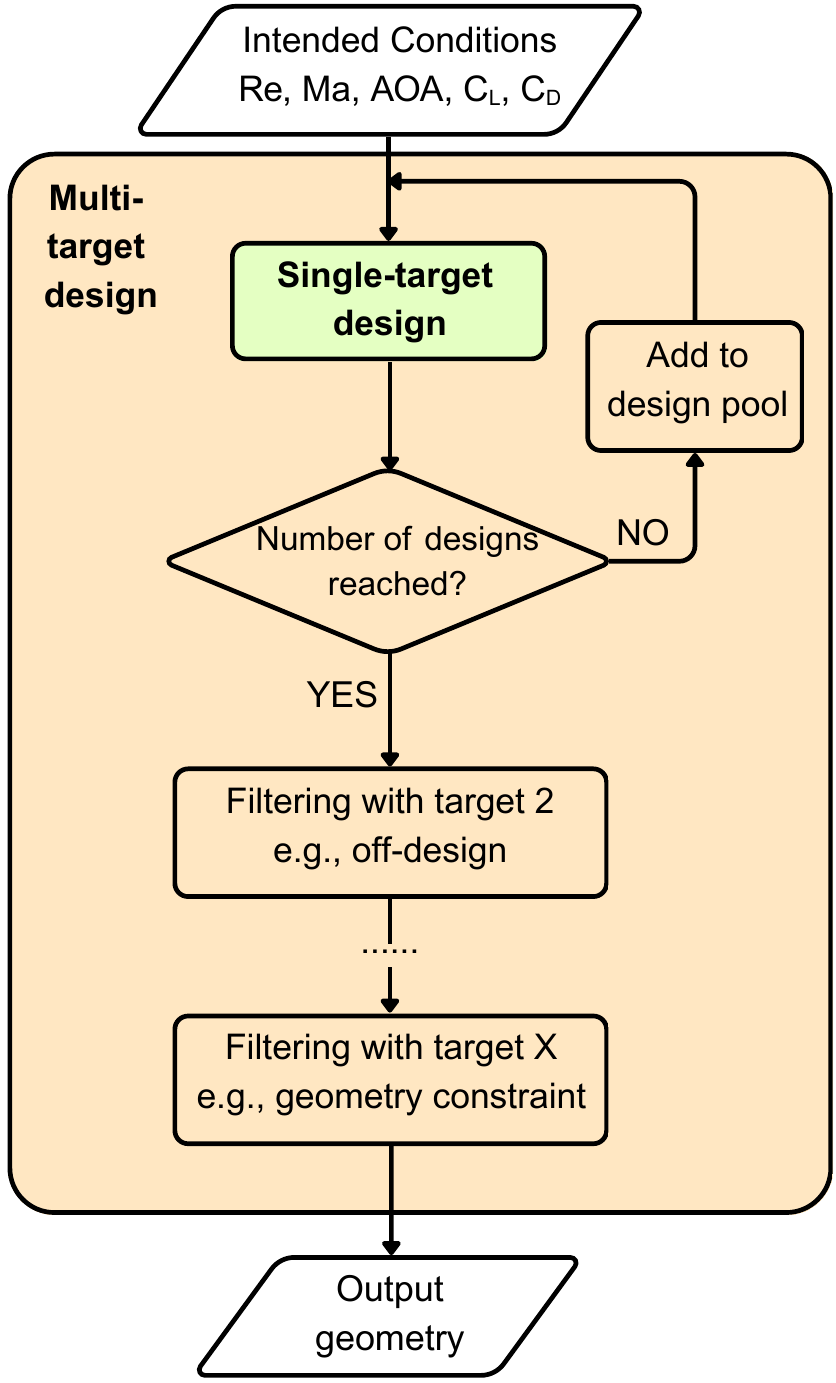}
    \caption{Multi-Target Generation}
    \label{multi target generation flow chart}
    \end{subfigure}
    \caption{Diffusion Model Generation Procedure}
    \label{generation flow charts}
\end{figure}

In this study, tolerances ($\varepsilon$) of $C_L$ and $C_D$ in Fig.~\ref{single-target generation flow chart} are set to be 1\,\% and 5\,\%, respectively. A maximum of 100 design trials is allowed; if the errors are still larger than the tolerance, the best performing geometry with the lowest RMSE error is selected. The choice of maximum 100 design trials is set for reference in this study and can be adjusted according to the specific needs in practice for better performance. The number of trials required to generate a design within the tolerance of reaching target $C_L$ and $C_D$ (100 if maximum design trials are reached) is recorded as a measure of the effectiveness of the generative model. In addition, another metric, $N_{95}$, is used to quantify the average number of trials needed before reach a probability of $95\,\%$ of obtaining at least one feasible design to satisfy the specified tolerances. It is defined in Eq.~\eqref{N95}. Here, $q$ is the cumulative probability of obtaining at least one successful design within $N$ independent trials, $p$ denotes the probability of a single generated sample satisfying the prescribed tolerances, and $N_{95}$ corresponds to the value of which $q = 95 \,\%$. 

\begin{equation}
    \label{N95}
    q = 1 - (1-p)^N
\end{equation}

It is also noticed that there are also cases where the XFOIL fails to converge with the generated airfoil profile. This is usually caused by the lack of smoothness of the generated profiles, and these designs are considered as physically unfeasible. A model failure rate is hence defined as the percentage of the testing set that can not generate any feasible designs after maximum design trials as a reflection of model robustness.

\subsection{Model Applications}
\subsubsection{Single-target generation}
The single-target generation procedure follows the workflow illustrated in Fig.~\ref{single-target generation flow chart}, with the aerodynamic target $C_L$ and $C_D$ being the only design objectives. In this study, the model's single-target design ability is investigated from several aspects, from interpolation to extrapolation. 

\paragraph{In-distribution} The model is tasked with the testing set samples, which are essentially an interpolated subset from the training set. This is the most standard way of diffusion model deployment, and is also used in this study to compare the model performances across the three different data formats. 

\paragraph{Continuous Condition Interpolation} To reduce data volume, the training set uses a coarse, discretized grid of flow conditions in Section.~\ref{Airfoil Database and Label Generation section}. Because the flow conditions are physically continuous, we evaluate the model's continuous interpolation ability by introducing a test set with off-grid flow conditions (i.e. intermediate values between the training levels of $Re$, $M$, and $AOA$), and compare performance against the in-grid baseline. To achieve this, the testing set samples flow conditions are perturbed randomly such that they are at intermediate values compared to the training labels (e.g., $Re = 280000$, $Ma = 0.15$, $AOA=2.5$).

\begin{figure}[h]
    \centering
    \includegraphics[width=0.7\textwidth]{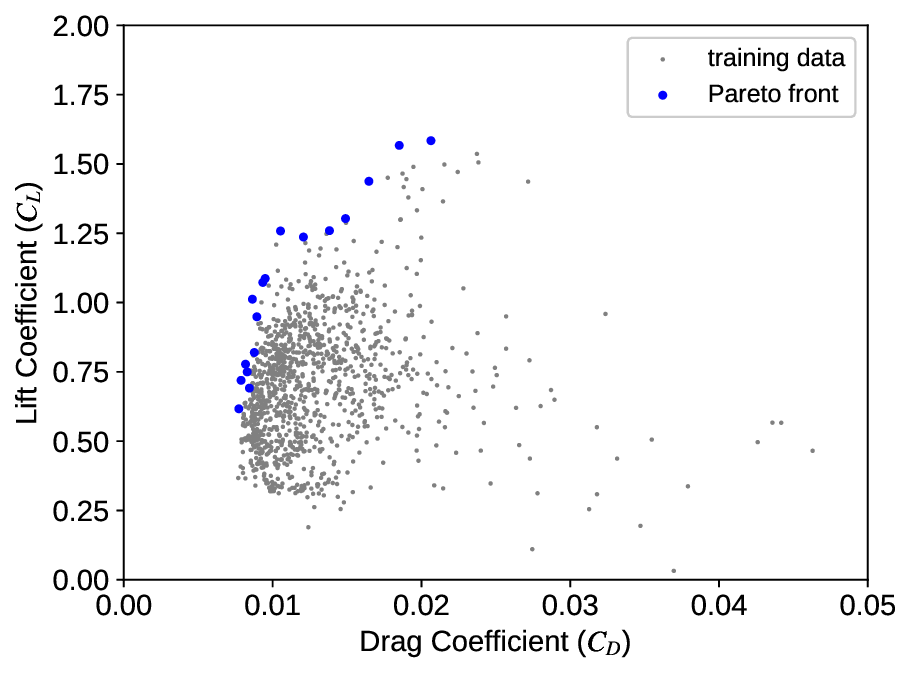}
    \caption{Distribution of the Testing Samples in an Example Flow Condition $Re = 250000$, $Ma = 0.2$, $AOA=3$. Pareto Front Marked in Blue. }
    \label{extrapolation distribution}
\end{figure}

\paragraph{Aerodynamic Extrapolation}
To test the ability of aerodynamic extrapolation ability, the model is tasked to generate airfoils with target $C_L$ values outside the boundary of the training dataset. Training dataset's Pareto front, which in this paper is defined as the best aerodynamically performing samples with the highest lift-to-drag ratio under a given flow condition, is first identified for each flow condition. An illustration of the Pareto front is shown in Fig.~\ref{extrapolation distribution} for $Re = 250000$, $Ma = 0.2$, $AOA=3$. The model is then used to generate airfoils that surpass the performance under the same flow condition ($Re$, $Ma$, and $AOA$) of the training set Pareto front by gradually increasing $C_L$ with the same $C_D$. 

\paragraph{Flow Condition Extrapolation}

To investigate the model's extrapolation ability, a testing set with flow condition values sampled outside the training dataset boundary. Four featuring extrapolated flow conditions are selected for this test, as listed in Table~\ref{flow condition extrapolation table}. 

\begin{table}[hbt!]
    \caption{Extrapolated Flow Conditions}
    \label{flow condition extrapolation table}
    \centering
    \begin{tabular}{ccccccc}
    \hline
    \textbf{Case}& \textbf{Extrapolated Quantity} & \textbf{Re}& \textbf{Ma}& \textbf{AOA} & $\mathbf{C_L}$ & $\mathbf{C_D}$\\
    \hline
    1 & $Re$ & 400000 & 0.6 & 5 & 0.55 & 0.04 \\
    2 & $AOA$ & 250000 & 0.5 & -1 & 0.5 & 0.03 \\
    3 & $Ma$ & 300000 & 0.7 & 3 & 0.5 & 0.03 \\
    4 & All & 400000 & 0.7 & -1 & 0.4 & 0.03 \\
    \hline
    \end{tabular}
\end{table}

\subsubsection{Multi-target Design}
As explained in Section~\ref{intro section}, the stochastic nature of the diffusion model is particularly advantageous for inverse design problems, as it enables the generation of diverse designs that satisfy the specified targets. This property allows for more effective exploration of the design space and the identification of multiple feasible solutions for a given set of conditions, making diffusion models especially suitable for multi-target design workflows. An example workflow of the multi-objective design generator is shown in Fig.~\ref{multi target generation flow chart}. By default, the first target corresponds to the target $C_L$ and $C_D$ at the given flow condition. Additional targets can be specified as needed, and can be geometric, (e.g., leading edge radius and maximum thickness) or off-design performance (e.g. stall limit, $C_L$ and $C_D$ at a different flow condition). As the number of targets or objectives increases and the tolerance for each target becomes tighter, more design trials are required to satisfy all objectives.

As shown by Fig.~\ref{multi target generation flow chart}, the single-objective generation procedure (Fig.~\ref{single-target generation flow chart}) is applied repeatedly to collect a series of designs that satisfy the $C_L$ and $C_D$ targets, after which the resulting design pool is filtered according to the additional objectives. Although this multi-objective design workflow involves repeated sampling, it is not a purely random search of the design space. Each generated geometry is conditioned on the target $C_L$ and $C_D$, leveraging the learned distribution of feasible designs. The stochastic nature of the diffusion model allows multiple distinct designs to satisfy the same aerodynamic targets, naturally capturing the one-to-many mapping of the inverse problem. Compared to the literature methods, which directly use multiple design targets as diffusion model condition inputs~\cite{Gong2024}, the post-processing of solution filtering in our proposed workflow presents a flexible choice of any arbitrary secondary targets. Once trained, the diffusion model can be used for any secondary targets without the need to retrain the diffusion model.

By combining repeated single-objective generation with filtering according to secondary objectives, the workflow performs an efficient, guided stochastic exploration of the design space, enabling diverse solutions that meet multiple targets simultaneously. In this study, the stall-point angle of attack is used as a secondary target to demonstrate the multi-target generation workflow, illustrating how diverse and feasible solutions can be efficiently explored using diffusion model.

\section{Results}

\subsection{Single-Target Generation Results} 
\subsubsection{In-distribution}
\label{direct interpolation section}
\begin{figure}[h]
    \centering
    \includegraphics[width=\linewidth]{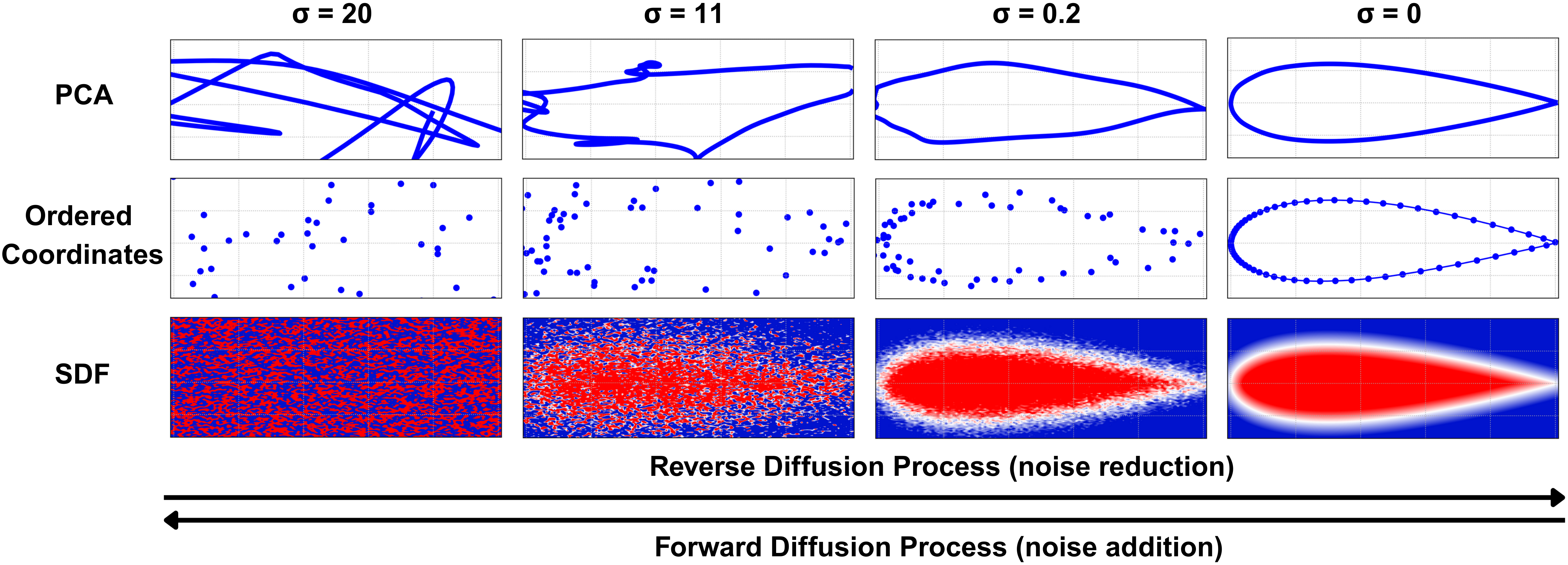}
    \caption{Denoising Process of Three Geometry Encoding Methods. $\sigma$ Represents the Noise Level}
    \label{denoising process}
\end{figure}

The sampling process of diffusion model with each data format is illustrated in Fig.~\ref{denoising process}. Four noise levels of $\sigma = 0, 0.2, 11$, and $20$ are displayed to present the forward and reverse diffusion process.
For the ordered coordinate- and SDF-based geometry representations, the denoising process can be directly observed in the physical design space, where initially noisy and irregular profiles (on the left in Fig.~\ref{denoising process}) are gradually denoised into smooth and physically meaningful airfoil boundaries (on the right in Fig.~\ref{denoising process}). In contrast, for the PCA-based representation, the diffusion process operates in the latent space of principal component weights as explained in Section~\ref{geometry representation subsubsection}. As a result, the denoising behavior is not directly visible in the geometry space in Fig.~\ref{denoising process} for PCA-based method, and sensible airfoil shapes only emerge after the denoised principal component weights are linearly combined to reconstruct the final geometry.

\begin{figure}[hbt!]
    \centering
        \centering
    \includegraphics[width=\textwidth]{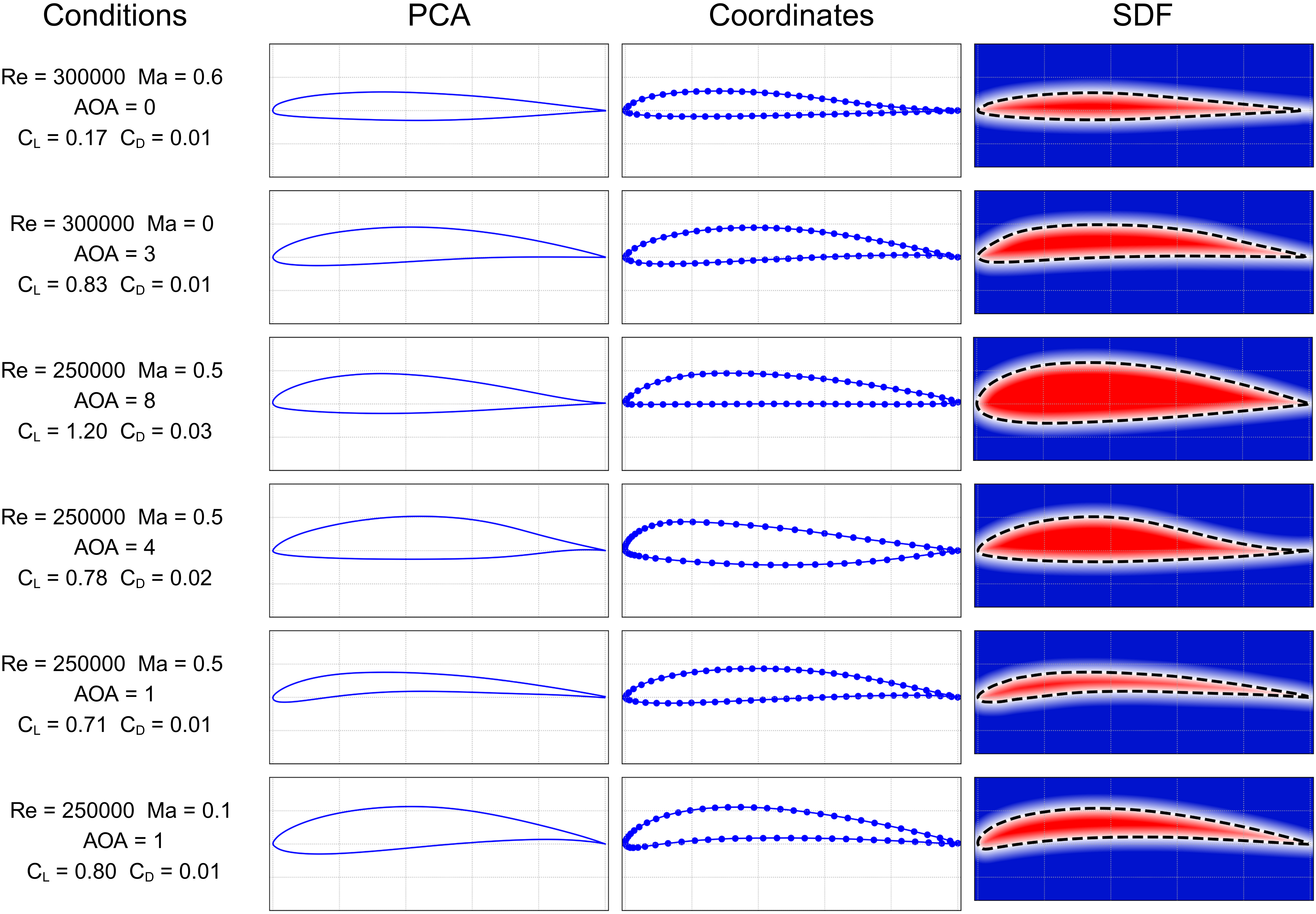}
 
    \caption{Examples of Generated Airfoil Profiles}
    \label{examples of generated profiles}
\end{figure}

Six example airfoil profiles generated using the single-target generation workflow (Fig.~\ref{single-target generation flow chart}) are shown in Fig.~\ref{examples of generated profiles} for comparison. All generated profiles are smooth and free from geometric artifacts, indicating that the diffusion models successfully capture the underlying shape manifold of the training data for all three data structures. Despite being conditioned on identical target aerodynamic coefficients, the generated profiles exhibit noticeable geometric differences across the three data representations. This diversity arises from the stochastic nature of the diffusion sampling process and reflects the inherent one-to-many mapping of the inverse design problem. Such variability is desirable, as it provides multiple feasible design candidates that satisfy the same performance targets, offering greater flexibility for downstream design selection and optimization.

\begin{figure}
    \centering
    \begin{subfigure}[t]{0.9\textwidth}
        \centering
        \includegraphics[width=\textwidth]{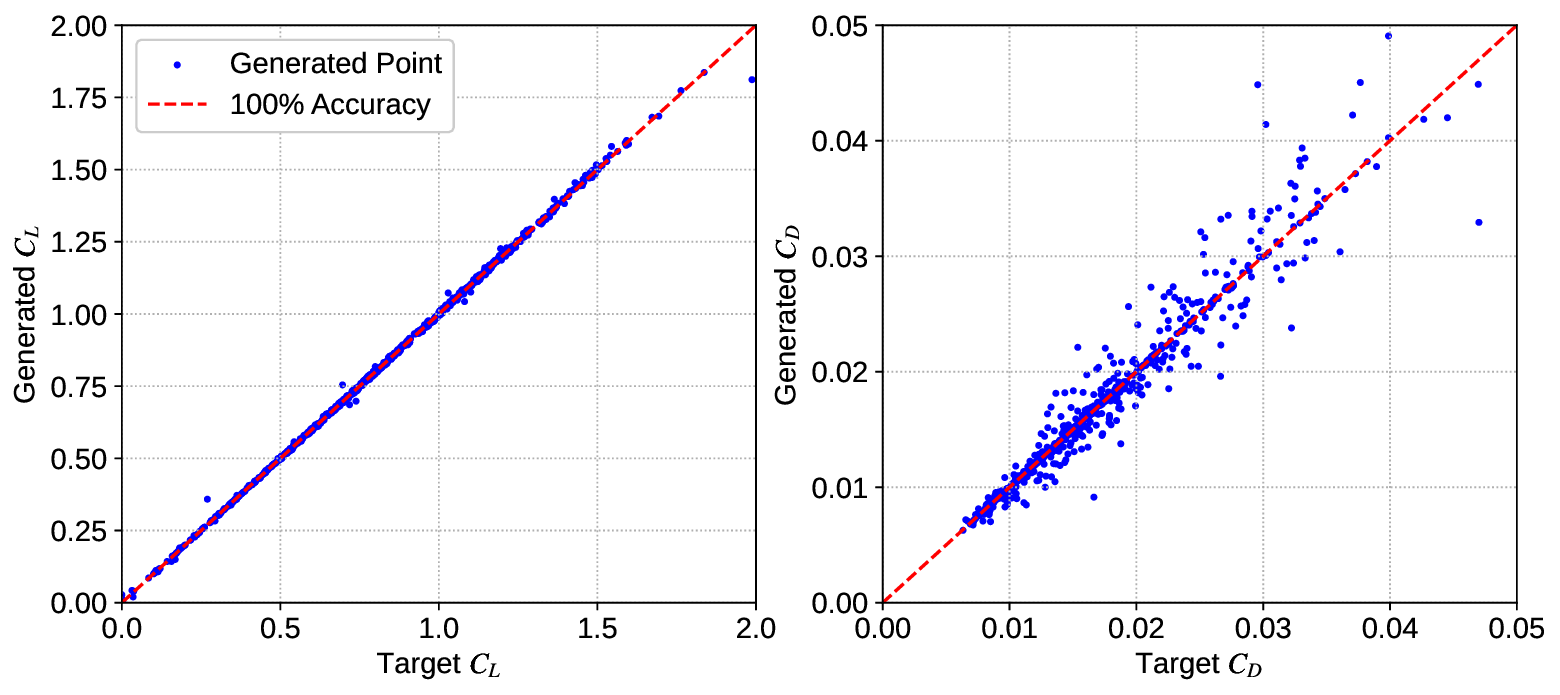}
        \caption{PCA}
        \label{PCA single-target}
    \end{subfigure}
    \hfill
    \begin{subfigure}[t]{0.9\textwidth}
        \centering
        \includegraphics[width=\textwidth]{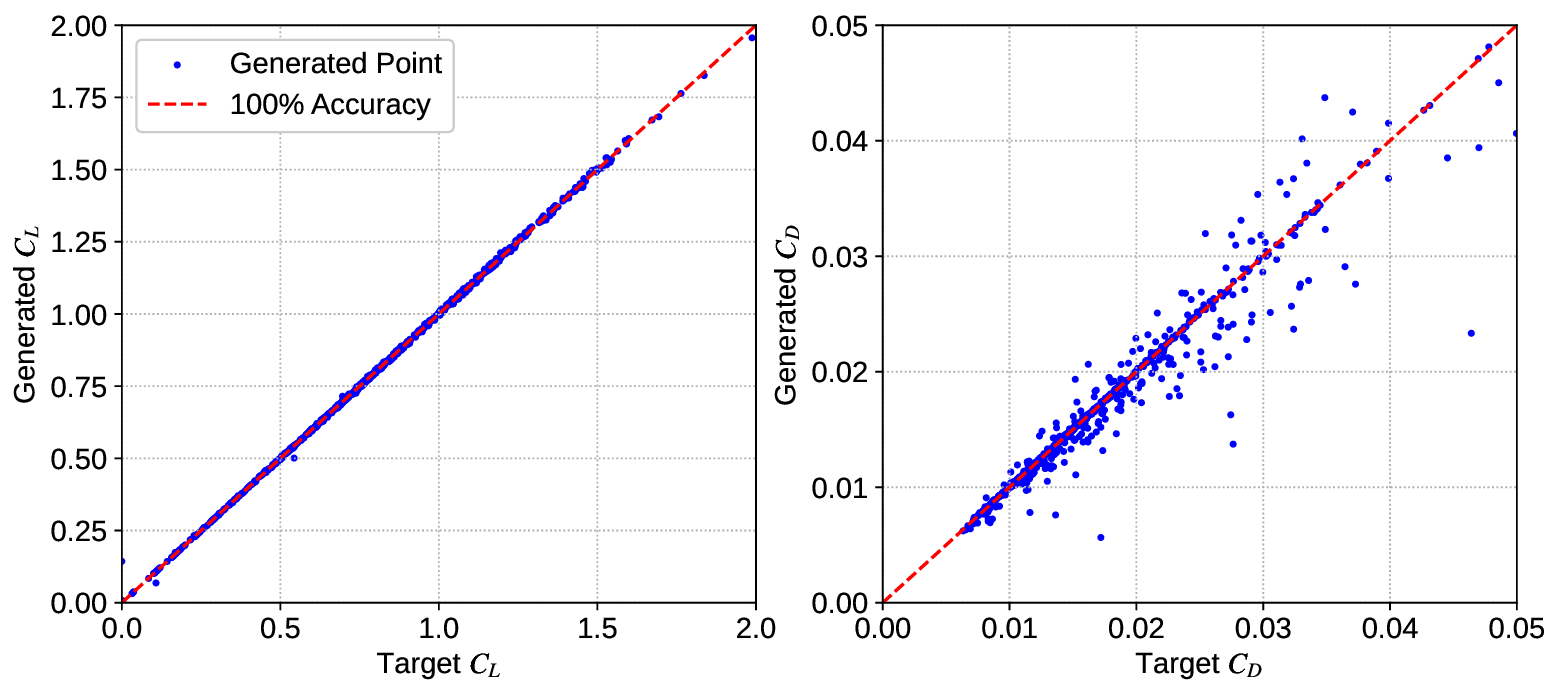}
        \caption{Coordinates}
        \label{coordinate single-target}
    \end{subfigure}
    \hfill
    \begin{subfigure}[t]{0.9\textwidth}
        \centering
        \includegraphics[width=\textwidth]{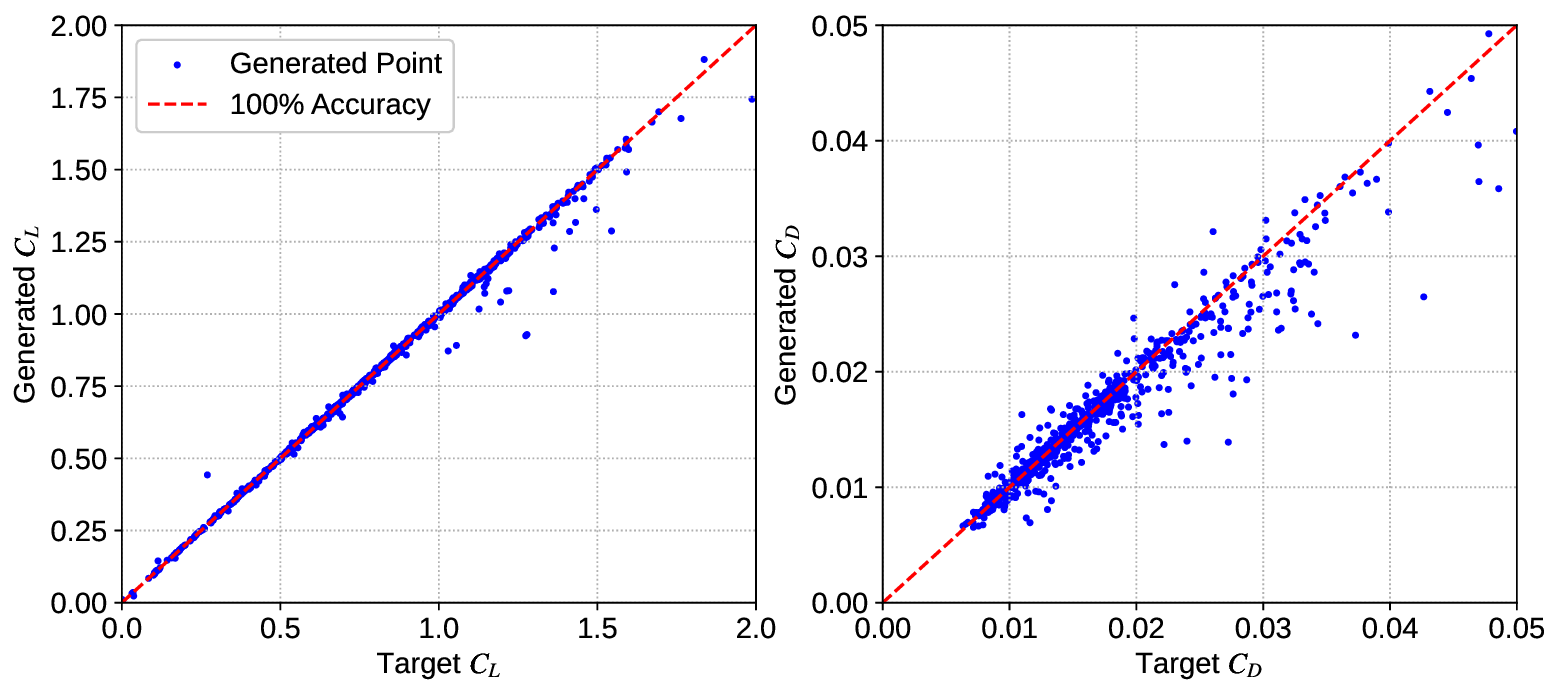}
        \caption{SDF}
        \label{SDF single-target}
    \end{subfigure}
    
    \caption{Single-target Generation Testing Set Results}
    \label{single-target generation results}
\end{figure}

\begin{table}[hbt!]
    \caption{Single-target Generation Results}
    \label{single-target generation results table}
    \centering
    \begin{tabular}{cccccc}
    \hline
    \textbf{Data Format} & \textbf{$C_L$ RMSE}& \textbf{$C_D$ RMSE}& \textbf{Average Trials} & \textbf{Unfeasible Designs} & \textbf{$N_{95}$} \\
    \hline
    PCA & 3.1\% & 8\% & 61 & 0.13\,\% & \textbf{29} \\
    Coordinates & \textbf{0.15\%} & \textbf{7.5\%} & 56 & \textbf{0\%} & 33\\
    SDF & 3.8\% & 9.5\% & \textbf{55} & 0.4\% & 97\\
    \hline
    \end{tabular}
\end{table}

To better visualize the model performance, the testing set results of single-target generation are presented in Fig.~\ref{single-target generation results}. The XFOIL-calculated aerodynamic metrics ($C_L$ and $C_D$) of generated airfoil profiles are plotted against the specified targets in Fig.~\ref{single-target generation results}. The $C_L$ and $C_D$ are presented on the left and right, respectively, and the zero-error line $y=x$ is shown with red dashed lines for reference. The performance metrics are then summarized in Table~\ref{single-target generation results table}. All three models are able to generate physically feasible airfoils within specified tolerances for the majority of the specified conditions.

It can be seen that, in the case of single-target generations, the diffusion model trained with ordered coordinates performs the best across all metrics. It achieves the target $C_L$ and $C_D$ values with the least number of trials, results in the lowest RMSEs, and generates no unfeasible profiles as shown in Table~\ref{single-target generation results table}. The PCA-based model performs slightly worse than the ordered coordinates model. SDF-based model exhibits the lowest performance metrics among the three data representations, yet it still achieves reasonable agreement, as shown in Fig.~\ref{single-target generation results}.

In terms of the model robustness of generating smooth, physically feasible airfoils, although in literature it is reported that latent space diffusion (PCA and CST) provides the advantages of smoothness guarantee compared to explicit diffusion (coordinates and mesh)~\cite{Wagenaar2024}, it is observed that PCA-based diffusion model still resulted in a small number of unfeasible designs, whereas for ordered coordinates there is no single occurrence of such case in our testing set. SDF results in the greatest amount of unfeasible designs. This is partially due to the insufficient training data explained in the previous paragraph, and may also because of the lower effective resolution of SDF compared to coordinates. 
The additional post-processing steps, including the quadratic interpolation and coordinates reordering, increases the chance of loss of smoothness, leading to XFOIL convergence failure and hence have the highest percentage of 0.67\,\% of unfeasible designs in Table~\ref{training_details_table}. It is also noticed that for all three geometry encoding approaches, the error associated with $C_D$ label is significantly larger than $C_L$. This is because the $C_D$ values are at an order of magnitude smaller than $C_L$ values, and hence it will be more sensitive to geometry changes than $C_L$.

\subsubsection{Continuous Condition Interpolation}
\begin{figure}
    \centering
    \begin{subfigure}[t]{0.9\textwidth}
        \centering
        \includegraphics[width=\textwidth]{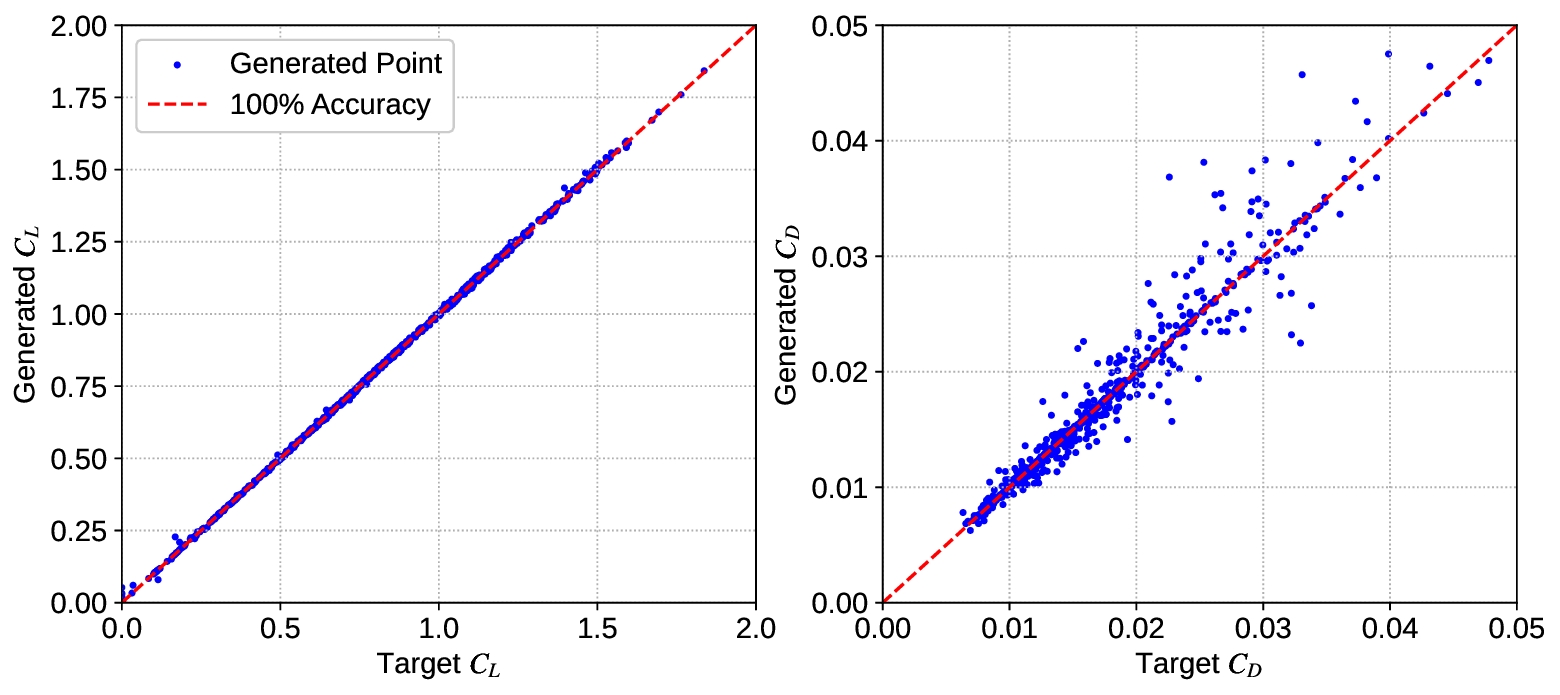}
        \caption{PCA}
        \label{PCA intermediate}
    \end{subfigure}
    \hfill
    \begin{subfigure}[t]{0.9\textwidth}
        \centering
        \includegraphics[width=\textwidth]{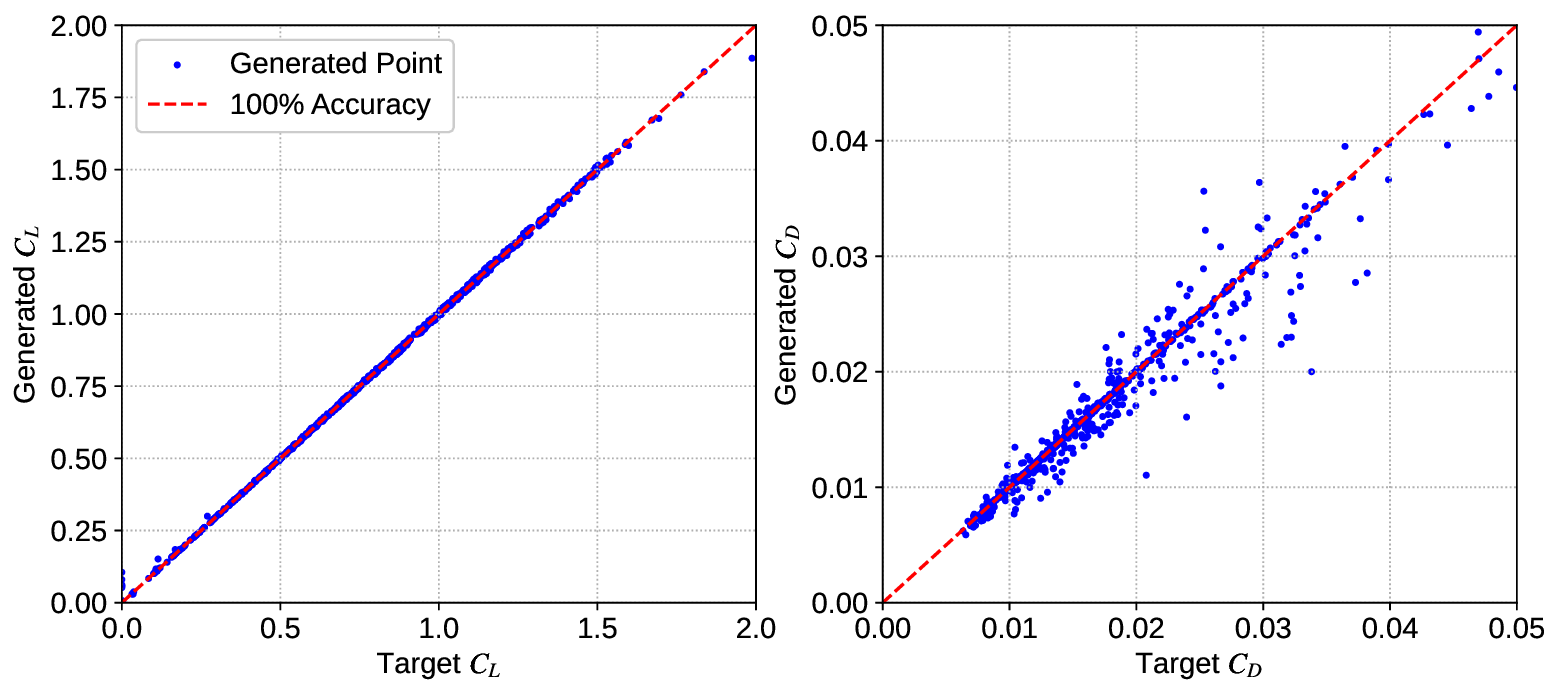}
        \caption{Coordinates}
        \label{coordinate intermediate}
    \end{subfigure}
    \hfill
    \begin{subfigure}[t]{0.9\textwidth}
        \centering
        \includegraphics[width=\textwidth]{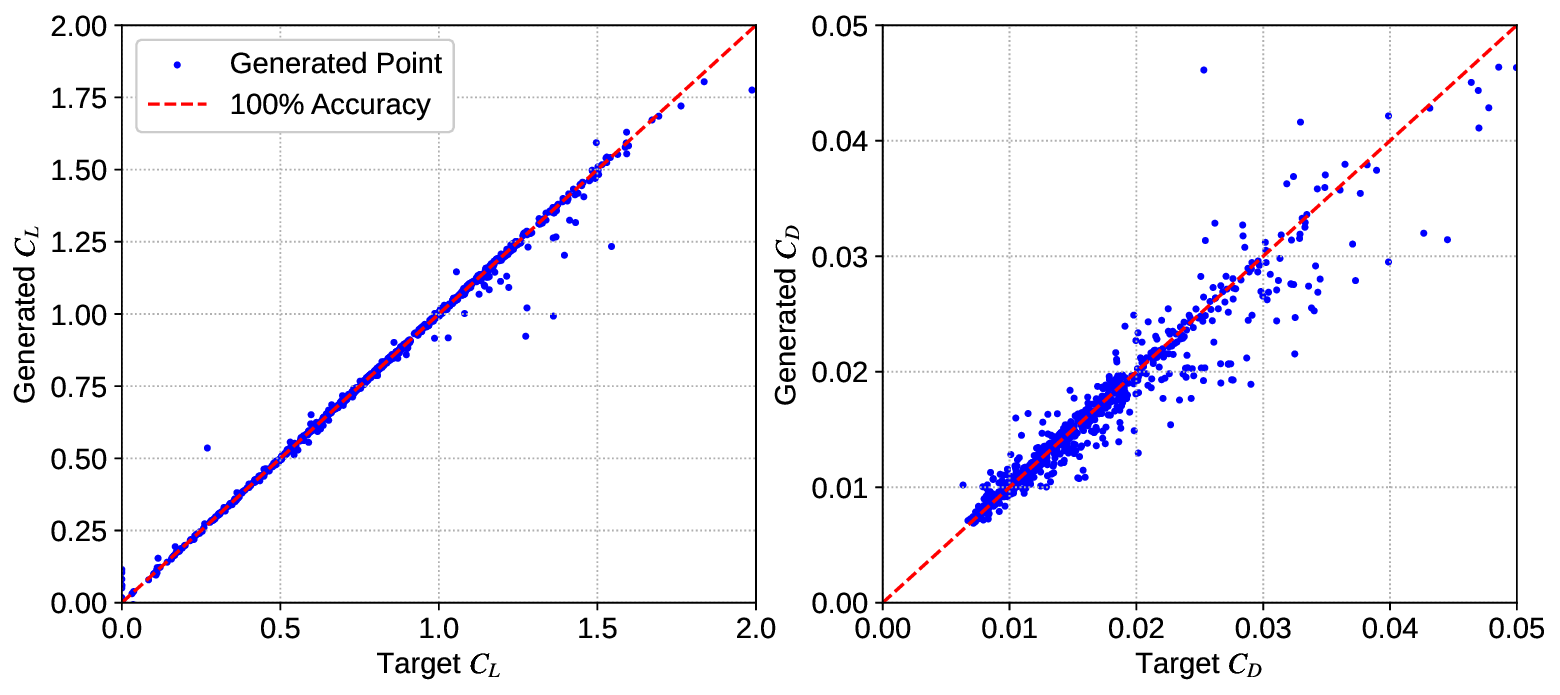}
        \caption{SDF}
        \label{SDF intermediate}
    \end{subfigure}
    
    \caption{Single-Target Generation with Intermediate Flow Conditions Results}
    \label{intermediate results}
\end{figure}

The results for testing set cases with continuously interpolated flow conditions are shown in Fig.~\ref{intermediate results} and in Table.~\ref{intermediate results table}. The results are similar to the ones in Section~\ref{direct interpolation section} -- the ordered coordinate-based model performs the best, followed by the PCA-based model. The SDF-based model shows the worst performance among three data structures again. 

Additionally, compared to the corresponding Table.~\ref{single-target generation results table}, in which all the flow conditions are the same as in the training set, the model retains the same level of performance on continuously interpolated flow conditions. This indicates that, despite the condition labels in the training set is discrete, the model still managed to capture the continuous relationship between geometry and the condition label well.

\begin{table}[hbt!]
    \caption{Single-Target Generation with Continuous Condition Interpolation Results}
    \label{intermediate results table}
    \centering
    \begin{tabular}{cccccc}
    \hline
    \textbf{Data Format} & \textbf{$C_L$ RMSE}& \textbf{$C_D$ RMSE}& \textbf{Average Trials} & \textbf{Unfeasible Designs} & \textbf{$N_{95}$}  \\
    \hline
    PCA & \textbf{3.2\%} & 8\% & 61 & 0.40\,\% & \textbf{32} \\
    Coordinates & 3.4\% & \textbf{7.3\%} & \textbf{56} & \textbf{0.00\%}  & 39\\
    SDF & 4.6\% & 10.5\% & 58 & 0.40\% & 98 \\
    \hline
    \end{tabular}
\end{table}

\subsubsection{Aerodynamic Extrapolation}
\begin{figure}[ht]
    \centering
    \includegraphics[width=0.8\textwidth]{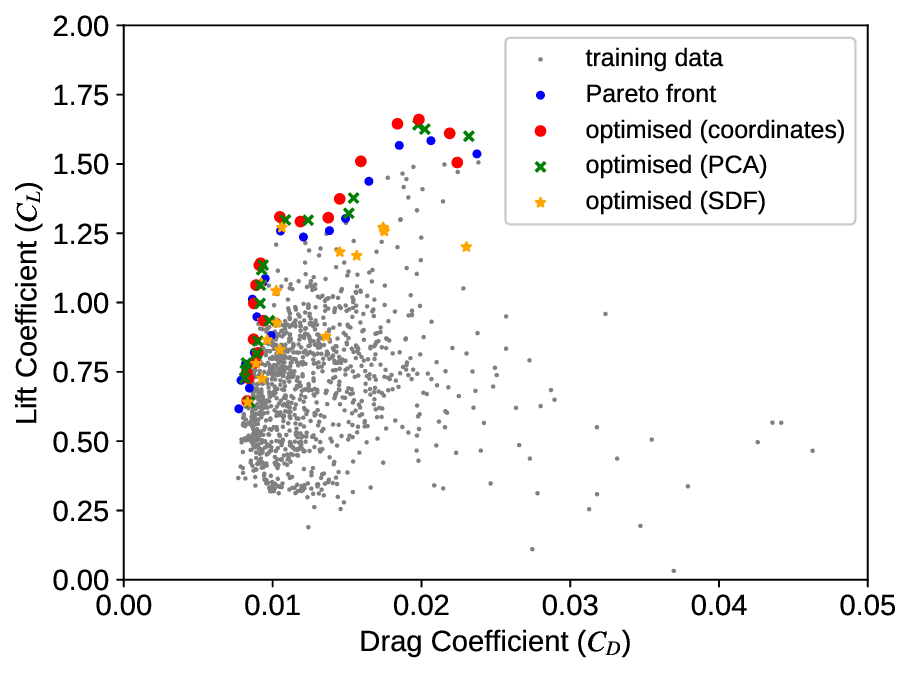}
    \caption{Aerodynamic Extrapolation Results}
    \label{optimize beyond pareto results}
\end{figure}

The aerodynamic extrapolation results using the single-target generation procedure are shown in Fig.~\ref{optimize beyond pareto results} and in Table~\ref{aerodynamic extrapolation results table}. A similar trend can be observed here again. The coordinates-based diffusion model achieves the greatest amount of successful optimizations with the greatest $C_L$ increments. The PCA-based diffusion model performs slightly worse. The SDF-based model completely failed in this aerodynamic extrapolation test -- not only did it fail to achieve a single case of successful optimization, but it also could not recover the original Pareto front of the training dataset. This is likely due to the insufficient training data and low effective resolution as explained previously. The low training data density at the Pareto front is insufficient to train the SDF-based model. 

These results proves the excellent extrapolation ability of the coordinate-based diffusion model beyond the aerodynamic limits of the training set. It is also noted that the optimization at the higher $C_D$ region is more likely to success compared to the low $C_D$ region. This could be due to the original Pareto front at the low-drag region already reaching the aerodynamic limits. Although the diffusion model shows the ability to extrapolate aerodynamically, it cannot generate airfoils that exceeds the limits of physics, and hence is unable to optimize these points further. 

\begin{table}[hbt!]
    \caption{Aerodynamic Extrapolation Results}
    \label{aerodynamic extrapolation results table}
    \centering
    \begin{tabular}{ccc}
    \hline
    \textbf{Data Format} & \textbf{No. Successful Optimization}& \textbf{Average $C_L$ Increment} \\
    \hline
    PCA & 6 & 3\,\%  \\
    Coordinates & \textbf{12} & \textbf{5\,\%} \\
    SDF & 0 & N/A \\
    \hline
    \end{tabular}
\end{table}


\subsubsection{Flow Condition Extrapolation Results}

\begin{table}[hbt!]
\caption{Extrapolated Flow Conditions Results}
\label{flow_condition_extrapolation_results_table}
\centering
\begin{tabular}{c c c c c c c}
\hline
 & \multicolumn{2}{c}{\textbf{PCA}}
 & \multicolumn{2}{c}{\textbf{Coordinates}}
 & \multicolumn{2}{c}{\textbf{SDF}} \\
\cline{2-3}\cline{4-5}\cline{6-7}
\textbf{Case}
 & \textbf{$C_L$ Error} & \textbf{$C_D$ Error}
 & \textbf{$C_L$ Error} & \textbf{$C_D$ Error}
 & \textbf{$C_L$ Error} & \textbf{$C_D$ Error} \\
\hline
1 & \textbf{3.49\,\%} & 9.42\,\% & 7.61\,\% & \textbf{3.81\,\%} & 8.16\,\% & 20.2\,\% \\
2 & 0.98\,\% & 4.33\,\% & \textbf{0.12\,\%} & \textbf{13.4}\,\% & 5.62\,\% & 13.7\,\% \\
3 & 1.16\,\% & 10.2\,\% & \textbf{1.01\,\%} & \textbf{0.62\,\%} & 0.5\,\% & 11.4\,\% \\
4 & 0.68\,\% & \textbf{3.33\,\%} & \textbf{0.20\,\%} & 13.8\,\% & 2.67\,\% & 25.6\,\% \\
\hline
\end{tabular}
\end{table}

The single-target generation results for the testing cases in the selected extrapolated flow condition  (Table.~\ref{flow condition extrapolation table}) are shown in Table.~\ref{flow_condition_extrapolation_results_table}. Overall, airfoils generated by the coordinate-based model achieves aerodynamic performance in best agreement with the targeted values, indicating the zero-shot extrapolation capability of the model.

\subsubsection{Discussion}

Overall, there is a clear trend that the diffusion model trained with ordered coordinates outperforms the other two models (PCA- and SDF-based models), showing superior performance across all performance evaluations, including aerodynamic performance error, average number of trials, unfeasible designs, interpolation, and extrapolation. The underlying reason that affects the model performance is the geometric information encoding process. 

For the ordered coordinates-based model, the geometric information is directly encoded and learned by the 1D convolutional neural layers of the diffusion model. The ordered arrangement of coordinates, as illustrated in Fig.~\ref{data formats}, preserves spatial continuity along the airfoil surface, enabling the convolution layers to efficiently extract the geometric features of the airfoil profiles.

In contrast, for the PCA-based diffusion, the geometric information is represented indirectly in latent space through the PCA transformation. The input of the diffusion model consists of a set of principal components weights, which do not retain explicit spatial information. As a result, the convolutional layers are less effective at exploiting their inherent advantage in local geometric feature extraction, making it more challenging for the model to learn the mapping from the condition labels to the corresponding airfoil geometries. 
In addition, as discussed in Section~\ref{intro section}, latent-space diffusion inherently reduces the effective design space through dimensionality compression. Although this can improve training efficiency, it may also limit the diversity of generated solutions. For inverse design problems, which are intrinsically one-to-many, this reduction in design space can be unfavorable when a wide variety of feasible designs is desired.

Among the three data representations, the 2D SDF-based diffusion model exhibits the lowest performance. This is primarily attributed to the increase in data dimensionality from one to two dimensions, which substantially raises the amount of trainable parameters, as shown in Table~\ref{training_details_table}, and consequently increases the amount of training data required. Therefore, for the same training dataset size, the SDF-based model underperforms compared to 1D geometry encodings like PCA and coordinates. In addition, the SDF representation relies on a discretized grid with finite resolution, which introduces a trade-off between geometric fidelity and computational cost. Limited resolution can lead to loss of fine geometric details and increased numerical diffusion during the denoising process.

Nevertheless, the promising results presented demonstrate the feasibility of using SDF to encode airfoil geometries for diffusion-based aerodynamic shape generation. As a continuous representation of geometry, the SDFs are well suited for representing high-dimensional geometric data in a continuous space, especially when the object can be described implicitly by a boundary, making it particularly suitable for complex geometric learning tasks such as 3D shape generation in engineering. The SDF work presented in this paper hence provides a clear pathway toward 3D shape generation, where higher-dimensional SDFs are particularly advantageous as a structured data format~\cite{Park2025}. 

\subsection{Multi-Targets Generation}
\begin{figure}[ht]
    \centering
    \begin{subfigure}[t]{\textwidth}
        \centering
        \includegraphics[width=0.6\textwidth]{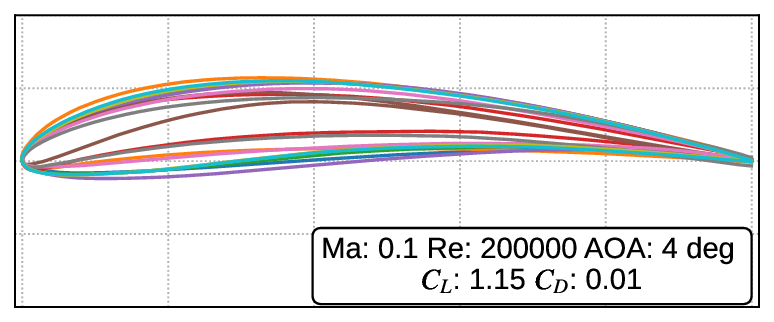}
        \caption{Feasible Design Pool}
        \label{design pool}
    \end{subfigure} 
        \hfill
    \begin{subfigure}[t]{0.47\textwidth}
        \centering
        \includegraphics[width=\textwidth]{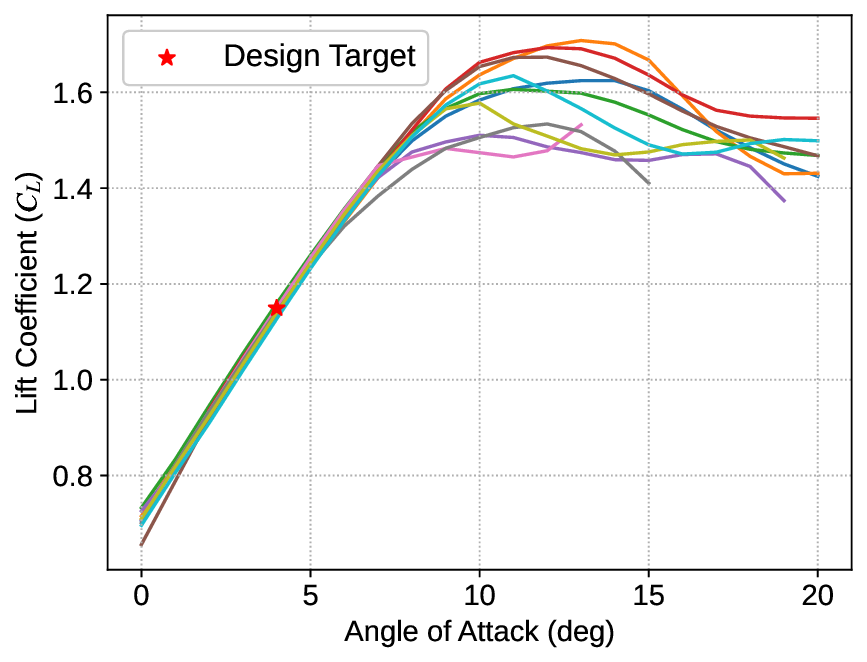}
        \caption{$C_L$ at Off-design AOAs}
        \label{CL off design}
    \end{subfigure}
        \hfill
    \begin{subfigure}[t]{0.49\textwidth}
        \centering
        \includegraphics[width=\textwidth]{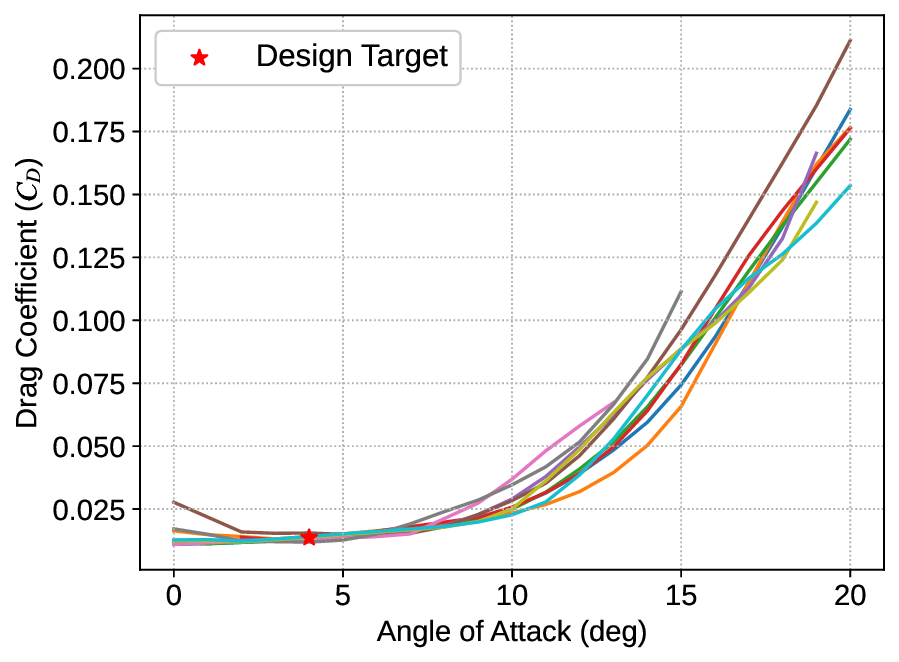}
        \caption{$C_D$ at Off-design AOAs}
        \label{CD off design}
    \end{subfigure}
    
    \caption{Multi-Targets Generation Results}
    \label{multi targets generation results}
\end{figure}

As discussed in Section~\ref{method model deployment and applications section}, the stochasticity of the diffusion model aligns with the nature of inverse design problems, which by definition is a one-to-many mapping. Therefore, there will be more than one possible geometry present which shares the same $C_L$ and $C_D$ values under the same flow conditions. Figure~\ref{design pool} shows an example of this non-deterministic mapping. To demonstrate the multi-targets generation workflow in Fig.~\ref{multi target generation flow chart}, ten different geometries are generated and all of them satisfy the $C_L$ and $C_D$ targets under the specified flow condition (labeled on the bottom right of Fig.~\ref{design pool}). By doing so, a design pool of ten airfoils that all satisfy the aerodynamic design target ($C_L$ and $C_D$) is formed. These designs can be further selected with different targets. The coordinate-based diffusion model, which shows the best performance in the single-target generation tasks, is used in this demonstration.

As an example of the multi-target generation, the off-design aerodynamic performance is evaluated by computing the $C_L$ and $C_D$ curves under the design flow conditions ($Ma$ and $Re$) but at off-design $AOA$s, as shown in Figs.~\ref{CL off design} and~\ref{CD off design}. The ten designs exhibit significantly different off-design performances, and the best performing profile can be selected according to the particular application requirement. For instance, if one requires the profile to have highest possible stall $AOA$, the orange profile will be selected. In fact, this selection can be applied to any targets, both geometrically and aerodynamically. 

Compared to the diffusion-model based multi-target generation approaches proposed in the literature, the presented method shows greater flexibility in real engineering applications. The multi-target generation paradigm proposed in the literature is to encode all the design targets as the condition label of the model~\cite{Diniz2024, Wen2026}, which may include maximum thickness, off-design performance, leading edge radius and so on. The generation process will therefore inherently become a one-shot multi-target generation process as shown in Fig.~\ref{single-target generation flow chart}. Although providing more controllability of the secondary design targets, this approach also reveals several drawbacks. The most obvious one is the need of preparing additional condition labels, which may include geometric labels and off-design performances. The additional conditions also increases the scale of the neural network needed and hence the training complexity. More importantly, the diffusion model trained with this approach will be case-specific and not universally applicable. If one wants to change the design targets (e.g., additional design requirements), the diffusion model will need to be re-trained with the new sets of design targets as the condition label. In contrast, despite the fact that our proposed method requires multiple design trials and post-filtering, it is also more flexible. 


\section{Conclusions}

In summary, a diffusion-based inverse-design framework for airfoils is introduced in this study, with $Re$, $Ma$ and $AOA$ as the flow conditions, and with $C_L$ and $C_D$ as the aerodynamic design targets. The model achieved accurate, constraint-satisfying generation across broad ranges of flow conditions, which sets the new state-of-the-art against the previous works in the literature. The conclusions and major contributions of this study are as follows. 

\begin{enumerate}

    \item The diffusion model is implemented as a novel paradigm of airfoil generation. By adopting the state-of-the-art EDM diffusion sampling method, jointly with multi-flow condition labeling of $Ma$, $Re$, and $AOA$ across broad ranges, this work presents an effective, unified diffusion model workflow to generate airfoil geometries that satisfy to the given design targets accurately. 

    \item This study provides the first systematic comparison of using different airfoil shape encoding data formats to train the diffusion model. Using the same diffusion backbone and the same U-Net architecture with ResNet blocks, three different encoding methods are implemented and compared: PCA-based method using PC weights, ordered $x$-$y$ coordinate, and SDF-based method using discretized 2D SDF values. The training effectiveness is evaluated from a wide range of perspectives, including both interpolation and extrapolation of the training dataset. The model's ability to generalize to interpolate at continuous flow condition labels is also tested by providing inputs that lie between the discrete values in the training set. The extrapolation ability is examined both aerodynamically, by tasking the diffusion model with $C_L$ target beyond the Pareto front of the training dataset, as well as with respect to flow conditions by tasking the model with flow conditions beyond the range of the training set.

    \item The comparison results consistently indicate that the model trained with coordinates outperform the PCA format on aerodynamic targets due to the explicit spatial encoding in its data structure, both with interpolation and extrapolation test sets. In addition, despite the SDF training data resulting in less performing model, it proves its feasibility in the field of aerodynamic shape generation, paving the way towards 3D shapes where SDF with more natural spatial encoding and scalability. 

    \item Based on the stochastic nature of the design generation with diffusion model, this study introduced multi-target design generation workflow using the trained diffusion model in engineering optimizations. In this paper, the diffusion model is deployed multiple times to collect a pool of designs that all satisfy the aerodynamic design target, from where the designs can be further filtered according to the particular application requirements.
\end{enumerate}

\bibliography{sample}

\end{document}